\documentclass[journal=jctcce,manuscript=article]{achemso}
\setkeys{acs}{usetitle = true}

\usepackage{amsmath}
\usepackage[american]{babel}
\usepackage{natbib}
\usepackage[utf8]{inputenc}
\usepackage{graphicx}
\usepackage{epsfig}
\usepackage{color}
\usepackage{subfigure}
\usepackage{dcolumn}
\usepackage{csquotes}
\usepackage{color}
\usepackage[normalem]{ulem}

\definecolor{gruen}{rgb}{0,0.4,0}
\newcommand{\mw}[1]{\textcolor{gruen}{\footnotesize MW: #1}}

\newcommand{\ch}[1]{#1}
\newcommand{\vek}[1]{\mathbf{#1}}
\newcommand{\bra}[1]{\langle #1|}
\newcommand{\ket}[1]{|#1\rangle}
\newcommand{\braket}[2]{\langle #1|#2\rangle}

\title{
  Ab-initio wave-length dependent Raman spectra: Placzek approximation and beyond. }

\author{Michael Walter}
\email{Michael.Walter@fmf.uni-freiburg.de}
\affiliation[IWM]{Fraunhofer IWM, MikroTribologie Centrum $\mu$TC, Wöhlerstrasse 11, D-79108 Freiburg, Germany}
\affiliation[FIT]{FIT Freiburg Centre for Interactive Materials and Bioinspired Technologies, Georges-K\"ohler-Allee 105, 79110 Freiburg, Germany}
\author{Michael Moseler}
\affiliation[IWM]{Fraunhofer IWM, MikroTribologie Centrum $\mu$TC, Wöhlerstrasse 11, D-79108 Freiburg, Germany}
\affiliation[IOP]{Physikalisches Institut, Universität Freiburg, Herrmann-Herder-Straße 3, D-79104 Freiburg, Germany}
\affiliation[FMF]{Freiburger Materialforschungszentrum, Universität Freiburg, Stefan-Meier-Straße 21, D-79104 Freiburg, Germany}

\begin{document}

\date{\today}

\begin{abstract}
We analyze how to obtain non-resonant and resonant
Raman spectra within the Placzek as well as the Albrecht
approximation.
Both approximations are derived 
from the matrix element for light scattering
by application of the Kramers, Heisenberg and Dirac formula.
It is shown that the Placzek expression results from a
semi-classical approximation
of the combined electronic and vibrational
transition energies.
Molecular hydrogen, water and butadiene are studied as test cases. 
It turns out that the Placzek approximation agrees qualitatively 
with the more accurate Albrecht formulation even in the resonant regime 
for the excitations of single vibrational quanta.
However, multiple vibrational excitations are absent in Placzek, 
but can be of similar intensities as single excitations under resonance conditions.
The Albrecht approximation takes multiple vibrational excitations into account 
and the resulting simulated spectra exhibit good agreement with experimental 
Raman spectra in the resonance region as well.  
\end{abstract}


\maketitle

\section{Introduction}

Raman spectroscopy is a facile and nondestructive tool
for the investigation of materials properties and
is applied in various fields of materials research.
The Raman effect couples light scattering to vibrational
modes and thus involves both electronic
and nuclear degrees of freedom as well as
the coupling thereof.
The complex properties of Raman spectra has entered
excellent textbooks \cite{Long02,Ferraro03}.

Despite the general availability in experiment, the theoretical
basis of the Raman effect is rather involved as it is a second
order process in the electronic degrees of freedom that
are coupled to nuclear vibrational degrees of freedom. 
Nevertheless, the calculation of Raman spectra by ab initio techniques 
has a long history. In the past,
most approaches are based on the powerful Placzek
approximation. In this approximation the Raman intensity is proportional
to the derivatives of the polarizability tensor with respect to
nuclear coordinates\cite{Porezag96prb,Long02}.
In chemistry related literature, the Placzek approximation is usually applied
only for excitation energies far from resonance 
and \ch{a large fraction} of approaches deduce the Raman
intensities from static polarizability 
derivatives.\cite{Yamakita07jcp,Castiglioni04,Rouille08,Zhao06nl,Porezag96prb,Martin15jmcc,Vecera17nat}
Some derivations go beyond this, but restrict
the calculation also to a single frequency far from
resonance \cite{Corni01jcpa,Cheeseman11,Barone14pccp,Bloino15jpca}. 
The Placzek approximation is frequently used 
in the resonance region in solid state Raman 
spectroscopy.\cite{Ambrosch02prb,Gillet13prb,Niu08pb,Wang18prb}
This has been criticized by some authors.\cite{Profeta01prb}

There are many approaches to address also resonant Raman spectra
that mostly start from the assumption that only Franck-Condon-type 
scattering is important\cite{Stock90jcp,Peticolas95jcc,Jarzecki01jrs,Neugebauer04jcp,Scholz11jcp,Balakrishnan12jpcb}.
Apart from pioneering early calculations\cite{Warshel77jcp},
only recently the application of an alternative formulation within
the Albrecht approximation \cite{Albrecht60jcp,Albrecht61jcp,Albrecht71jcp,Myers08}
has become tractable within standard electronic structure theory
\cite{Gong15jctc,Guthmuller16jcp,Heller16an,Duan16jctc,Hu16cms}.
In these approaches often only the resonant part is taken into
account \cite{Ferrer13jctc,Baiardi15jctc,Heller15,Guthmuller16jcp}
or the calculation is restricted to the static limit \cite{Gong15jctc}.

In the present work we study the calculation of 
Raman spectra of small molecules
in order to benchmark the different approaches
used in the literature and to elucidate their connection.
In particular, a formalism is presented that yields the Placzek as well as the 
Albrecht approximations from the Heisenberg-Kramers-Dirac matrix element. 
In this way the connection between the two approximations can be worked out explicitly.

This article is organized as follows. 
The underlying theoretical approach 
is detailed in the following section. 
Section \ref{sec:methods} describes the computational approach and 
the following sections reports the differences and similarities
of the different levels of theory for small molecules.

\section{Theory}

Raman scattering is nonelastic light scattering, where a
system of initial energy $E_I$ absorbs a photon
of energy $\hbar\omega_L$ with polarization $\vek{u}_L$ 
and ends up in
a state of energy $E_F$ having emitted a photon
with energy $\hbar\omega_S$ and polarization $\vek{u}_S$.
The cross section for this process can be expressed as
\cite{Albrecht61jcp,Porezag96prb,Scholz11jcp,Guthmuller16jcp}
\begin{equation}
  \label{eq:dsigma}
  \frac{d\sigma}{d\Omega} = 
  \frac{\omega_L\omega_S^3}{(4\pi\varepsilon_0)^2c^4}
  |V_{FI}|^2 \delta(E_I+\hbar\omega_L-E_F-\hbar\omega_S)
\end{equation}
where the $\delta$-function ensures energy conservation.
The second order matrix element for light scattering $V_{FI}$ 
has been derived by Kramers, Heisenberg and Dirac
and can be written as \cite{Kramers25zp,Dirac27prsla,Breit32rmp,
  Jensen05,Myers08}
\begin{equation}
  V_{FI}=\vek{u}_L\cdot\sum_{K\ne I}
  \left[\frac{\bra{I}\vek{D}\ket{K}\bra{K}\vek{D}\ket{F}}
    {E_K-E_I-\hbar\omega_L} +
    \frac{\bra{K}\vek{D}\ket{F}\bra{I}\vek{D}\ket{K}}
         {E_K-E_I+\hbar\omega_S}
         \right]\cdot\vek{u}_S \; ,
  \label{eq:VFI}
\end{equation}
where $\vek{D}$ denotes the dipole operator, a vector with the unit of 
length times charge.
Initial and final states are denoted by $\ket{I}$ and $\ket{F}$, respectively,
and include nuclear as well as electronic degrees of freedom.
The sum extends over all intermediate states $\ket{K}$ of the system
with their energies $E_K$.
Often an imaginary part is added to the photon energies
$\hbar\omega_L$ and $\hbar\omega_S$
in eq. (\ref{eq:VFI}). This does not affect our further derivations,
however.

In order to make the calculation tractable,
the Born-Oppenheimer approximation has to be applied resulting in 
the separation of electronic and nuclear degrees of freedom.
For Raman scattering, the electronic ground state is adopted both in 
the initial and the final states.
We therefore write the initial state $\ket{I}$ ($I\equiv 0,i^0$) 
and final state 
$\ket{F}$ ($F\equiv 0,f^0$) 
as products of the electronic ground state $\ket{0}$ and
the corresponding initial $\ket{i^0}$ and final $\ket{f^0}$
vibrational states.   
The intermediate states $\ket{K}$ ($K\equiv e, k^e$)
are products of excited electronic states $\ket{e}$ and the corresponding
vibrational states $\ket{k^e}$.
Furthermore, we assume that the light is exclusively absorbed by the
electronic system\ch{, i.e. no photon absorption by the nuclear wave 
function is considered.} This should be a good approximation
for the usual laser frequencies in the optical range 
\ch{utilized in Raman spectroscopy}.
The dipole transition matrix elements expressed in these 
states can be written as 
\begin{equation}
  \bra{I}\vek{D}\ket{K} = \bra{i^0}  \vek{m}_e(\xi)\ket{k^e} \; , \; \;
  \vek{m}_e(\xi)=\bra{0(\xi)}\vek{D}\ket{e(\xi)} \; ,
\end{equation}
where the dependence of the electronic states
on the nuclear coordinates $\xi$ is made explicit.
Using the electronic energies of ground and excited
states $E_0$ and $E_e$, and the corresponding  
vibrational energies 
$\varepsilon^e_k$ and $\varepsilon^0_i$,  we
may express the matrix element as
\begin{eqnarray}
    V_{FI}=\vek{u}_L\cdot\sum_{e}\sum_{k}
  \left[\frac{\bra{i^0}\vek{m}_e\ket{k^e}\bra{k^e}\vek{m}^*_e\ket{f^0}}
  {E_e-E_0+\varepsilon^e_k-\varepsilon^0_i- \hbar\omega_L}
  + 
    \frac{\bra{k^e}\vek{m}^*_e\ket{f^0}\bra{i^0}\vek{m}_e\ket{k^e}}
         {E_0-E_e+\varepsilon^e_k-\varepsilon^0_i+\hbar\omega_S}
  \right]\cdot\vek{u}_S \; .
  \label{eq:KHDBO}
\end{eqnarray}
\ch{Despite that the energies $E_0$ and $E_e$ correspond to the
  electronic system they are not depending on nuclear coordinates at this
  point. Within the independent mode double harmonic (IMDHO)\cite{Neese07ccr}
  approximation the energies
  represent the minima of the of the harmonic
  oscillators shown in Fig.~\ref{fig:DisplacedHO} below.
}

\mw{A:} Note, that the matrix element $V_{FI}$ itself does not select initial
and final vibrational states $i^0$ and $f^0$, respectively.
State selection is enforced by energy conservation in eq.
(\ref{eq:dsigma}).  The energy difference between the 
laser and the emitted photon \ch{directly reflects} the 
\ch{energy difference of} vibrational states \ch{as}
\begin{equation}
  \hbar\omega_L-\hbar\omega_S=E_F-E_I=\varepsilon_f^0-\varepsilon_i^0 \; .
\end{equation}
This difference is positive for Stokes and negative for Anti-Stokes scattering.

\begin{figure}[h]
  \centering{
    \includegraphics[width=0.8\textwidth]{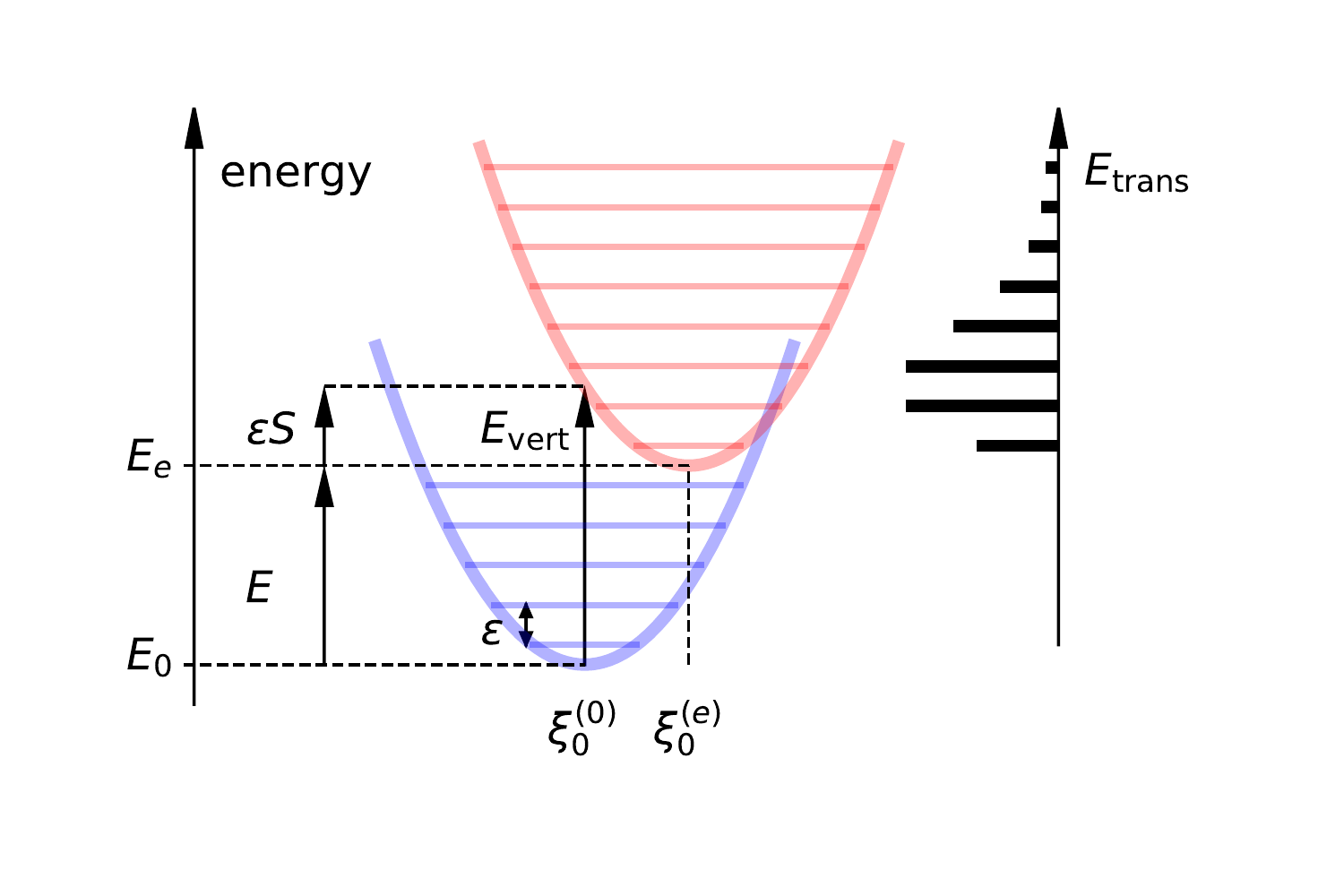}}
  \caption{Displaced harmonic oscillator model for the electronic 
    ground state (blue)
    and one excited state (red). The black bars on the right 
    show the size of the squared Franck-Condon factors for a transition
    from the vibrational ground state.}
  \label{fig:DisplacedHO}
\end{figure}
Starting from (\ref{eq:KHDBO}) the widely used 
Placzek approximation is obtained by employing the
semi-classical approximation\ch{\cite{Lee83jcp}} to replace the \ch{energy}
terms $E_e-E_0+\varepsilon^e_k-\varepsilon^0_i$
by \ch{their main contribution that is}
the vertical transition energy to this specific
electronic state, i.e.
\begin{equation}
  \label{eq:semiclassical}
  E_e-E_0+\varepsilon^e_k-\varepsilon^0_i\approx E_{{\rm vert}, e} \;,
\end{equation}
see Fig.~\ref{fig:DisplacedHO}.
\ch{The vertical transition energy depends on the nuclear position and
  is understood to be inside of the matrix element involving the initial
  vibrational state $\ket{i^0}$.}
The equivalent application of a classical Wigner phase space 
approximation\cite{Lee83jcp,Jensen05jcp}
leads to the same result.
We will discuss the validity of this assumption further below.
Applying the approximation (\ref{eq:semiclassical})
the $k$-dependence of the denominator in eq. (\ref{eq:KHDBO}) vanishes, and
the closure relation within the vibrational subspace
can be used\cite{Profeta01prb}
\begin{eqnarray}
  \sum_{k} \ket{k^e}\bra{k^e} = I_{{\rm vib}, e}
  \label{eq:completeness}
\end{eqnarray}
leading to
\begin{eqnarray}
  V_{FI}=\bra{i^0}\vek{u}_L\cdot\sum_{e}
  \left[\frac{\vek{m}_e\vek{m}^*_e}
    {E_{{\rm vert}, e}- \hbar\omega_L} +
    \frac{\vek{m}^*_e\vek{m}_e}
         {E_{{\rm vert}, e}+\hbar\omega_S}
  \right]\cdot\vek{u}_S \ket{f^0}.
  \label{eq:KHDsemi}
\end{eqnarray}
In case of real valued wave functions (as can always be assumed the case 
of finite systems in absence of magnetic fields)
and using $\omega_L\approx\omega_S$ in the denominator,
this matrix element further reduces to
\begin{equation}
  V_{FI}= \vek{u}_L\cdot\bra{i^0}\alpha(\omega_L) \ket{f^0}\cdot\vek{u}_S
  \label{eq:Valph}
\end{equation}
where the polarizability tensor
\begin{equation}
  \label{eq:alpha}
  \alpha(\omega)=\sum_e\frac{2E_{{\rm vert}, e} \vek{m}_e\vek{m}_e}
  {E_{{\rm vert}, e}^2-\hbar^2\omega^2} \; . 
\end{equation}
of the system in its electronic ground state emerges.

Eq. (\ref{eq:Valph}) has important consequences as 
it represents the overlap between initial and final
vibronic states in the electronic ground state
that are orthonormal.
In order to get a non-vanishing vibrational contribution (apart from the
Rayleigh scattering, where $f^0=i^0$),
an operator depending on vibrational coordinates is
required.
This dependence can be extracted by expanding
$\alpha(\omega)$ in terms of normal vibrational 
coordinates\cite{Porezag96prb} $Q_v$
around the nuclear equilibrium position $\xi_0$, where usually
only the first order is taken into account
\begin{equation}
  \alpha(\omega;\xi)=\alpha(\omega;\xi_0)
  + \sum_v \left.\frac{\partial \alpha(\omega;\xi_0)}{\partial Q_v}
  \right|_{Q_v=0}Q_v \; .
  \label{eq:dalpha}
\end{equation}
The first term in the expansion (\ref{eq:dalpha})
corresponds to Rayleigh scattering and the second term
contributes to the Raman effect giving 
\begin{equation}
  |V_{FI}|^2=\sum_v |\bra{i^0}Q_v\ket{f^0}|^2 
  \left|\vek{u}_L\cdot\frac{\partial \alpha(\omega_L)}{\partial Q_v}
    \cdot\vek{u}_S\right|^2 \; .
  \label{eq:0Q1alpha}
\end{equation}
The orthogonality of the vibrational states in the electronic ground state
shows that coupling to vibrational excitations
is due to the first derivatives in eq. (\ref{eq:dalpha}) only, and
that only single vibrational quanta can be introduced
by light scattering within this approximation.
\ch{Anharmonic effects ore mode mixing\cite{Montero82jcp,Barone14pccp,Bloino15jpca}
  might lead to multiple vibrational excitations. These effects
  are of second order in the derivative after vibrational coordinates
  (c.f. Fig. \ref{H2} below) and thus beyond the IMDHO model applied here.
}

The Placzek approximation is very successful for \ch{molecules}.
These have a large electronic ``gap'', such that the usual experimental
excitation wavelengths $\omega_L$ 
in the infrared or visible regions
are far from any electronic resonances of the molecules.
Many calculations even assume the limit $\omega_L\to 0$ and calculate
$I_{\rm Ram,v}$ from the static polarizability derived
by calculations with static electric fields 
\cite{Hemert81mp,Stirling96jcp,Porezag96prb,Shinohara98jms,Jackson97prb,Umari05drm,Niu08pb,Vecera17nat}.
This intensity is often interpreted as ``the'' Raman intensity
although experimental approaches report a wavelength
dependence of Raman spectra since decades \cite{Myers08}.

The assumption that all electronic resonances are far from $\hbar\omega_L$
is not valid anymore in solids. 
Raman spectra are the primary source of information
to characterize amorphous carbon for example, where the usual $\omega_L$
are well in the range of electronic excitation frequencies.\cite{Ferrari00prb}
Therefore strong effects from variations in $\omega_L$
are reported.\cite{Piscanec05drm}
This motivated Profeta and Mauri \cite{Profeta01prb} to 
express eq. (\ref{eq:KHDsemi}) as
function of two independent sets of nuclear coordinates $\xi, \xi'$
in $\vek{m}_e$ and $E_{{\rm vert}, e}$, respectively.
\ch{The polarizability tensor reads then
\begin{equation}
    \label{eq:alphaProfeta}
  \alpha(\omega;\xi,\xi')=\sum_e\frac{2E_{{\rm vert}, e}(\xi') \vek{m}_e(\xi)\vek{m}_e(\xi)}
  {E_{{\rm vert}, e}^2(\xi')-\hbar^2\omega^2} \; . 
\end{equation}}
The authors give some
reasoning to consider the derivatives after $\xi$ in $\vek{m}_e$, only,
which obviously is only part of the contribution.
We will call this contribution ``Profeta'' \ch{that is
  \begin{equation}
    |V_{FI}^{\rm Profeta}|^2=\sum_v |\bra{i^0}Q_v\ket{f^0}|^2 
    \left|\vek{u}_L\cdot
    \frac{\partial \alpha(\omega_L;\xi,\xi')}{\partial Q_v}
    \cdot\vek{u}_S\right|^2 \;
  \label{eq:0Q1alphaProfeta}
  \end{equation}
  with the vibrational coordinates $Q_v$ corresponding to the
  nuclear coordinates $\xi$. This part is labeled
  ``three-band terms'' by Wang et al.\cite{Wang18prb}
  We name the remaining part
} of (\ref{eq:KHDsemi}) ``Pl/Pr'' as shorthand for
Placzek without Profeta in the following.
\ch{It explicitly reads
   \begin{equation}
    |V_{FI}^{\rm Pl/Pr}|^2=\sum_v |\bra{i^0}Q_v\ket{f^0}|^2 
    \left|\vek{u}_L\cdot
    \frac{\partial \alpha(\omega_L;\xi,\xi')}{\partial Q_v'}
    \cdot\vek{u}_S\right|^2 \;
  \label{eq:0Q1alphaPlPr}
  \end{equation}
  with the vibrational coordinates $Q_v'$ corresponding to the
  nuclear coordinates $\xi'$.
  This part was labeled as ``two-band terms'' in Wang et al.\cite{Wang18prb}
}
We will see further below that neglecting the ``Pl/Pr''
contribution can be a severe and misleading approximation
\ch{at least in molecular systems} as it
disregards the resonant part of the Raman contributions.

In order to go beyond the Placzek approximation,
one may start form eq. (\ref{eq:KHDBO}) where we note
that in contrast to $E_{{\rm vert}, e}=E_{{\rm vert}, e}(\xi)$,
all energies are independent of
nuclear coordinates.
We may expand already
the matrix elements in terms of normal
coordinates\cite{Albrecht61jcp,Rousseau76,Guthmuller16jcp}
\begin{eqnarray}
  \vek{m}_e(\xi)=\vek{m}_e(\xi_0) + 
  \sum_v \left.\frac{\partial \vek{m}_e}{\partial Q_v}
  \right|_{Q_v=0}Q_v \; .
  \label{eq:dme}
\end{eqnarray}
The first term of this expansion
leads to the Albrecht A term
\begin{eqnarray}
    V_{FI}^A=\sum_{e}\sum_{k}
    \vek{u}_L\cdot\left[\frac{\vek{m}_e\braket{i^0}{k^e}
        \braket{k^e}{f^0}\vek{m}_e}
    {E_e-E_0+\varepsilon^e_k-\varepsilon^0_i- \hbar\omega_L}+
    \frac{\vek{m}_e\braket{i^0}{k^e}\braket{k^e}{f^0}\vek{m}_e}
         {E_0-E_e+\varepsilon^e_k-\varepsilon^0_i+\hbar\omega_S}
  \right]\cdot\vek{u}_S
  \label{eq:AlbrechtA}
\end{eqnarray}
where we use the shorthand notation $\vek{m}_e=\vek{m}_e(\xi_0)$.
It is also called the Franck-Condon term\cite{Dierksen04jcp,Guthmuller16jcp} and
is believed to be dominating for $\omega_L$ in resonance
with optically strong transitions\cite{Scholz11jcp}.
Note, that a non-negligible contribution from Albrecht A
to the off resonant Raman spectrum of water
was reported recently.\cite{Gong15jctc}
Close to resonance often
only the first part inside of the brackets is kept since this 
is the dominating contribution due to the small denominator.
The other part is non-resonant
and hence much smaller, such that
\begin{eqnarray}
  \label{eq:vFIres}
  V^A_{FI}=\sum_{e}\vek{u}_L\cdot\vek{m}_e\vek{m}_e\cdot\vek{u}_S
  \sum_{k}
  \frac{\braket{i^0}{k^e}\braket{k^e}{f^0}}
  {E_e-E_0+\varepsilon^e_k-\varepsilon^0_i- \hbar\omega_L}
\end{eqnarray}
represents a good approximation in the neighborhood of 
resonances\cite{Moran02jcp,Jarzecki09jpca,Wachtler12,Baiardi15jctc}.
Eq. (\ref{eq:vFIres}) also shows that in case of a single, isolated resonance, 
as it is often present
in organic chromophors, the Raman cross section is
mainly determined by the weighted Franck-Condon overlaps
$\braket{i^0}{k^e}\braket{k^e}{f^0}$
corresponding to this single transition.\cite{Scholz11jcp}

\ch{Albrecht\cite{Albrecht61jcp} splits the contributions
  of}
the first derivatives in (\ref{eq:dme})
into a resonant part ($B$ term)
\begin{equation}
  V_{FI}^B=\sum_{e}\sum_{k}
    \vek{u}_L\cdot\frac{\braket{i^0}{k^e}
      \bra{k^e}Q_v\ket{f^0}\vek{m}_e \vek{m}^{v*}_e +
     \bra{i^0}Q_v\ket{k^e}\braket{k^e}{f^0} \vek{m}^{v}_e \vek{m}^*_e}
      {E_e-E_0+\varepsilon^e_k-\varepsilon^0_i- \hbar\omega_L}
      \cdot\vek{u}_S
  \label{eq:AlbrechtB}
\end{equation}
and a non-resonant part ($C$ term)
\begin{equation}
  V_{FI}^C=\sum_{e}\sum_{k}
    \vek{u}_L\cdot\frac{\braket{i^0}{k^e}
      \bra{k^e}Q_v\ket{f^0}\vek{m}^{v*}_e \vek{m}_e +
      \bra{i^0}Q_v\ket{k^e}\braket{k^e}{f^0}\vek{m}^*_e \vek{m}^{v}_e
    }
         {E_e-E_0+\varepsilon^e_k-\varepsilon^0_i+ \hbar\omega_S}
      \cdot\vek{u}_S                               
      \; ,
  \label{eq:AlbrechtC}
\end{equation}
where the shorthand $\vek{m}^v_e=\partial\vek{m}_e/\partial Q_v$ is used.
The sum of these two terms are labeled Albrecht $B$ term by
Gong et al\cite{Gong15jctc} and is also called
Franck-Condon/Herzberg-Teller term\cite{Dierksen04jcp,Guthmuller16jcp}.
There is also the possibility to consider both derivatives
in the matrix elements. This so called Herzberg-Teller
term\cite{Guthmuller16jcp} is believed to be only important
when the matrix elements vanish, i.e. for symmetry forbidden
transitions \cite{Korenowski78jcp} and is not considered in our work.

In order to simplify the dependence on the polarization vectors $\vek{u}_L$ 
and $\vek{u}_S$,
the so called Raman invariants\cite{Long02} can be defined from the
tensor elements of $\alpha'=\partial \alpha/\partial Q_v$ for the Placzek 
approximation.
These are the mean polarizability\cite{Woodward49tfs,Porezag96prb,Long02}
\begin{equation}
  a = \frac{1}{3}(\alpha_{xx}' + \alpha_{yy}' + \alpha_{yy}')\, ,
  \label{eq:a}
\end{equation}
the anisotropy\cite{Long02,Guthmuller16jcp} (this quantity is also denoted by 
$\beta$\cite{Porezag96prb,Jackson97prb} or $g$\cite{Baiardi15jctc})
\begin{align}
  \gamma^2 = \frac{1}{2}\left[|\alpha_{xx}'-\alpha_{yy}'|^2 +
    |\alpha_{xx}'-\alpha_{zz}'|^2 + |\alpha_{yy}'-\alpha_{zz}'|^2\right] +
  \nonumber\\ 
  \frac{3}{4}\left[|\alpha_{xy}'+\alpha_{yx}'|^2 +
    |\alpha_{xz}'+\alpha_{zx}'|^2 + |\alpha_{yz}'+\alpha_{zy}'|^2\right]
\end{align}
and the asymmetric anisotropy\cite{Long02}
(often assumed to vanish\cite{Porezag96prb,Jackson97prb}
as expected for non-resonant Raman\cite{Ferraro03},
and also denoted by $d$\cite{Baiardi15jctc})
\begin{equation}
  \delta^2 = \frac{3}{4}\left[|\alpha_{xy}'-\alpha_{yx}'|^2 +
    |\alpha_{xz}'-\alpha_{zx}'|^2 + |\alpha_{yz}'-\alpha_{zy}'|^2\right]
  \label{eq:d2}
\end{equation}
from which the absolute Raman
intensity\cite{Porezag96prb,Jackson97prb,Niu08pb,Baiardi15jctc}
\begin{equation}
  I_{\rm Ram,v} = 45a^2 + 7\gamma^2 + 5\delta^2
  \label{eq:Iram}
\end{equation}
is obtained.
This intensity is usually given in units of
\AA$^4/$amu\ch{.}\cite{Porezag96prb}
\ch{Expression (\ref{eq:Iram}) is valid
only for the most common experimental setup and other 
combinations of $a, \gamma, \delta$ appear depending on the 
polarization of incoming and outgoing photons.\cite{Long02}}
Similar Raman invariants can also be defined from the tensor element of the
matrix element (\ref{eq:KHDBO}) instead of
\ch{the polarizability derivatives} $\alpha'$.\cite{Long02}
The resulting \ch{expression for the}
intensity is similar to (\ref{eq:Iram}) and is called $I$ in the 
following. \ch{In order to directly compare $I$ and $I_{\rm Ram,v}$
one would have to multiply the latter with the vibrational 
matrix element (Franck-Condon factor) $|\bra{i^0}Q_v\ket{f^0}|^2$.
The intensity $I$ is therefore} usually given in (e\AA/eV)$^2$.

In the following, we will compare Placzek and Albrecht approximations
using ab-initio calculations of small molecules. This will show
the similarities and differences of the two approximations.
We will find that the Albrecht and its semi-classical 
approximation Placzek largely agree for all excitation frequencies
and that in particular the approximation of Profeta corresponds to
Albrecht B/C and the missing terms Pl/Pr to Albrecht A.
The interested reader is also referred to Appendix \ref{app:AlbrechtPlaczek} 
that elaborates on a  
clear connection between Placzek and Albrecht in the limit 
$\omega_L\to 0$.

\section{Methods}
\label{sec:methods}

The electronic structure of the systems considered here
is described by density functional theory (DFT) as implemented in the 
GPAW software suite \cite{Mortensen05prb,Enkovaara10jpc}. 
The Kohn-Sham orbitals and the electronic density are described in 
the projector augmented wave (PAW) method \cite{Blochl94prb} where the 
smooth wave functions are represented on real space grids. 
The exchange correlation functional is approximated 
in the generalized gradient correction as devised by 
Perdew, Burke and Ernzerhof (PBE) \cite{Perdew96prl}. 
The real space grid was ensured to contain at least 4 \AA{} of vacuum space 
around each atom. 
The grid spacing for the wave-functions was chosen to be 0.2 \AA{}, 
while the density was represented \ch{on grids with 0.1 \AA{} spacing}. 
Molecular structures were considered to be relaxed when no 
force exceed 0.01 eV/\AA{}. 
Vibrational modes and frequencies are calculated
within the finite difference approximation of the
dynamical matrix\cite{Porezag96prb,Larsen17jpc}.
Excited state properties are calculated in time dependent 
DFT (TDDFT) linear response formalism as reported by Casida
\cite{Casida12arpc,Walter08jcp}.
\ch{The range of Kohn-Sham single-particle excitations was large enough
  to cover all the excitations in the energy ranges displayed,
  which also ensures convergence of the sum over states
  in polarizability derivatives and Alrecht terms.
  The Franck-Condon factors are calculated as described by Guthmuller.
  \cite{Guthmuller16jcp}
} 

We use the \ch{IMDHO} approximation that considers only
changes in excited state energies in linear order and no
mixing of ground state vibrational modes, i.e.
Duschinsky effects\cite{Guthmuller16jcp}
are thus not included.
Derivatives of \ch{polarizabilities, transition energies and} matrix elements
are calculated using finite differences.
Note, that this involves arbitrary phases related to the Berry phase
\ch{in case of the matrix elements from 
eq. (\ref{eq:dme}) and therefore} needs
special care as discussed in appendix \ref{app:Berry}.

\section{Results}

Molecular hydrogen is the simplest existing neutral molecule and
serves as a good example to show the basic properties and
consequences of the different
approximations for obtaining Raman intensities described above.
There is only one vibrational mode in H$_2$ 
which is found in our calculation at 4337 cm$^{-1}$ = 0.538 eV 
in fair agreement to the exact value of 
4163.3 cm$^{-1}$.\cite{Oklopcic16} 

\begin{figure}[h]
  \centering{
    \includegraphics[width=0.6\textwidth]{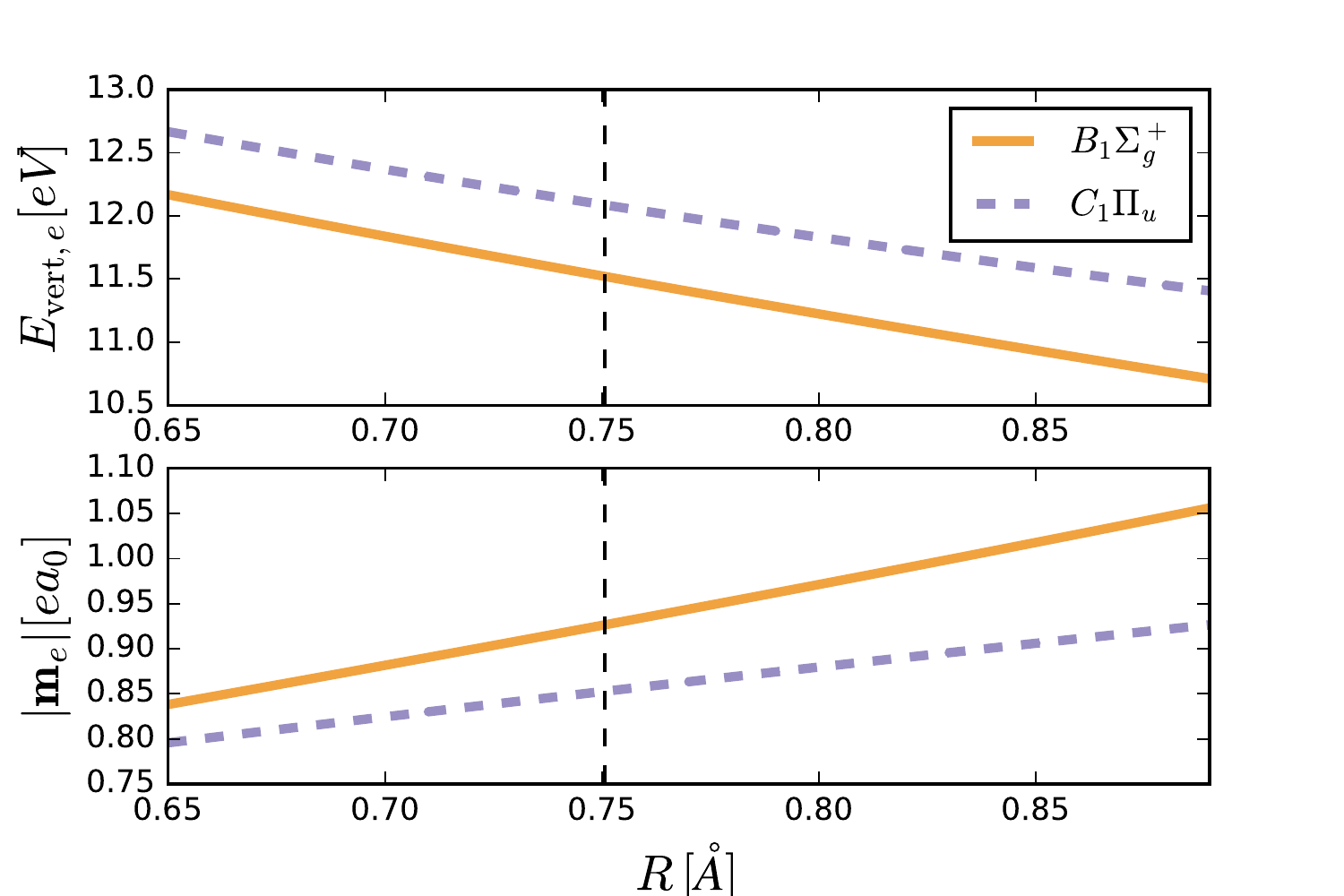}}
  \caption{Electronic transition energies $E_{{\rm vert}, e}$ and
    transition dipoles from the ground
    $^1\Sigma_u^+$ state to $B^1\Sigma_g^+, C^1\Pi_u$ states
    of the H$_2$ molecule depending on the nuclear distance $R$. 
    Vertical lines indicate the equilibrium distance.}
  \label{H2}
\end{figure}
The main ingredients determining Raman intensities are 
the derivatives of transition energy and dipole 
matrix element with respect to the normal coordinate of the 
corresponding vibration in Eqs. (\ref{eq:dalpha}) and (\ref{eq:dme}).
Fig. \ref{H2} shows these quantities
for the first two optically allowed transitions in the H$_2$ 
molecule in dependence of the bond length $R$.
Both quantities depend linearly on $R$ in agreement with the literature
\cite{Wolniewicz03,Fantz06}.
Adding a linear function to a Harmonic potential does
only change the potential minimum, but not its form,
i.e. the vibrational
frequencies in ground and excited state 
are the same (see also appendix \ref{appFC}).
Therefore the displaced harmonic oscillator model
\ch{underlying the definition of the Huang-Rhys parameter} \cite{Jong15pccp} is
indeed justified here.
Taking into account only the linear term in a
Taylor expansion of the energy around the equilibrium position
in the normal modes leads to the
displaced harmonic oscillator
model depicted in Fig. \ref{fig:DisplacedHO}.

\begin{figure}[h]
    \includegraphics[width=\textwidth]{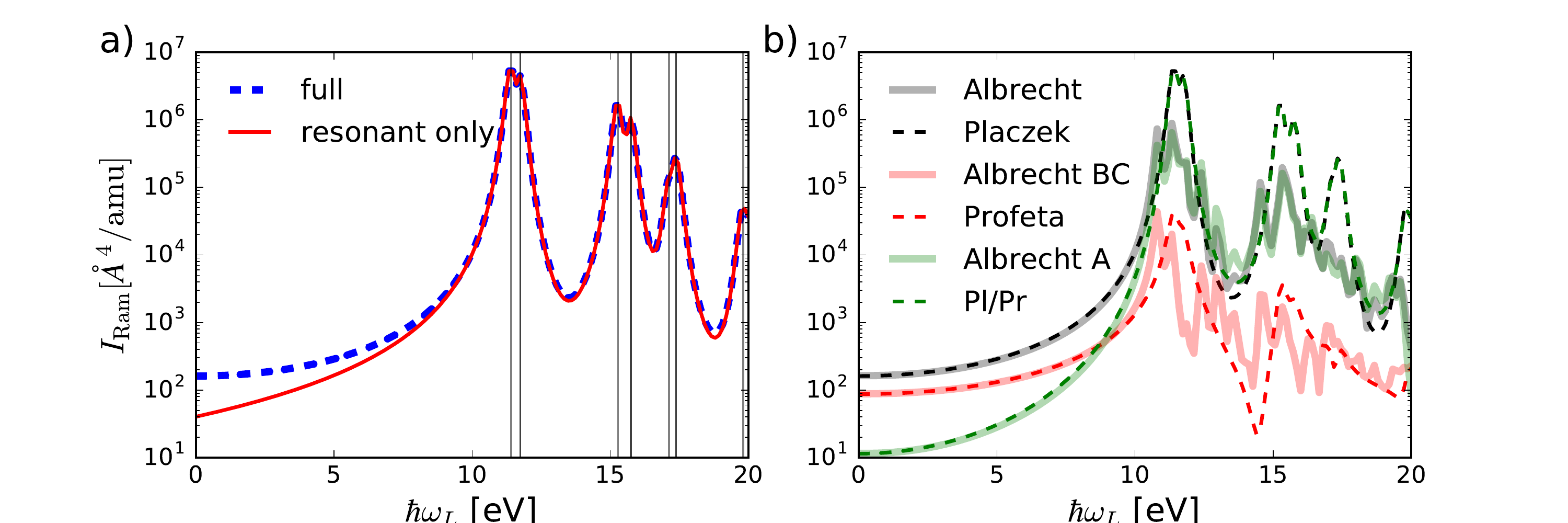}
  \caption{
    a) Full Placzek Raman intensity for H$_2$ compared to
    the resonant term only. Vertical lines show optically active
    transitions with oscillator strength $f\ge 0.01$.
    b)Raman intensity from different approximations.
    A width of $\gamma=0.2$ eV
    is applied for Placzek like approximations and $\gamma=0.1$ eV for
    Albrecht terms.
  }
  \label{H2raman}
\end{figure}
Fig. \ref{H2raman} a) shows the absolute Raman intensity 
in logarithmic scale as a function of the excitation frequency $\omega_L$.
The Raman intensity display a rather smooth dependence on photon
energy for small $\hbar\omega_L$, i.e. if $\hbar\omega_L\ll E_e$
for all optically strong electronic transitions energies $E_e$ in the system.
In contrast, the intensity gets strongly peaked close to the
transition energies 
if there is sufficient oscillator strength in the corresponding
transitions.
A finite width $\gamma$ is added as complex energy to
\ch{the full range of}
$\hbar\omega_L$ \ch{effectively broadening} these peaks.
Fig. \ref{H2raman} a) also compares the contributions
of the resonant and non-resonant terms in eq. (\ref{eq:KHDsemi}).
The non-resonant part can indeed be neglected, except when approaching
the static limit $\hbar\omega_L\to 0$, where the non-resonant part
is needed to get full intensity.

A comparison of the different approximations and their contributions for
H$_2$ is depicted in Fig. \ref{H2raman} b). 
We report the absolute Raman intensity although the
Albrecht terms do not contain the factor
$|\bra{i^0}Q_v\ket{f^0}|^2$ in eq. (\ref{eq:0Q1alpha}).
The Albrecht matrix elements have been divided by this factor to
get comparable numbers.
Concentrating
on the full Albrecht and Placzek approximation first, 
the similarity and even overlap of the two approximations 
for small $\omega_L$ far from the resonances becomes apparent. 
The two approximations yield the same result
in a wide energy range
and there is even a qualitative similarity
in the resonance regions.
The main difference is
that the less approximate Albrecht approximation leads to many more peaks
than Placzek.
The Placzek peaks are at the semi-classical vertical
transition energies where the denominator in eq. (\ref{eq:KHDsemi})
diverges. In contrast, the Albrecht terms exhibit peaks at each
of the phonon decorated electronic excitations in the
denominators of A, B and C terms in Eqs.
(\ref{eq:AlbrechtA}-\ref{eq:AlbrechtC}).
For this reason we had to apply twice the broadening $\gamma$
in the narrower peaks of Placzek as compared to Albrecht.

\begin{table}
  \begin{tabular}{l|r|r|r|r|r}
     & full & Albrecht A or Pl/Pr & Albrecht BC or Profeta \\
    \hline
    Albrecht & 191 & 11.5 & 109 \\
    Placzek  & 188 & 11.4 & 107
  \end{tabular}
  \\
  \caption{\label{tab:H2static}
    Static ($\omega_L=0$) absolute Raman intensities for the H$_2$ molecule
    in \AA$^4/$amu.
    }
\end{table}
Interestingly, and in agreement with the considerations of
appendix \ref{app:AlbrechtPlaczek}, the Profeta approximation
turns out to be the semi-classical approximation of the Albrecht BC terms.
The term missing in Profeta treatment corresponds to the Albrecht A term.
The connections and agreement between Albrecht and Placzek, and
Albrecht BC and Profeta are further corroborated by the numerical
values for static Raman intensities listed in tab. \ref{tab:H2static}. 
As expected, the Albrecht BC terms dominate for small $\omega_L$,
but these terms are not enough to give the full intensity.
Even in the limit
$\hbar\omega_L/E_e\to 0$ the consideration
of the Albrecht A contribution is important and cannot be 
neglected.
Albrecht A clearly dominates in the resonance region and
is the main contribution to the full Raman intensity.
In certain energy regions
the Albrecht A intensity is even larger than full Albrecht, which
indicates destructive interference with the Albrecht BC terms.
We note that there is no energy region where Albrecht BC
(and Profeta) is sufficient to give the correct intensities.
Placzek generally provides a good coarse grained description of the
intensity behavior as compared to Albrecht, however.

\begin{figure}[h]
  \centering{
    \includegraphics[width=0.6\textwidth]
    {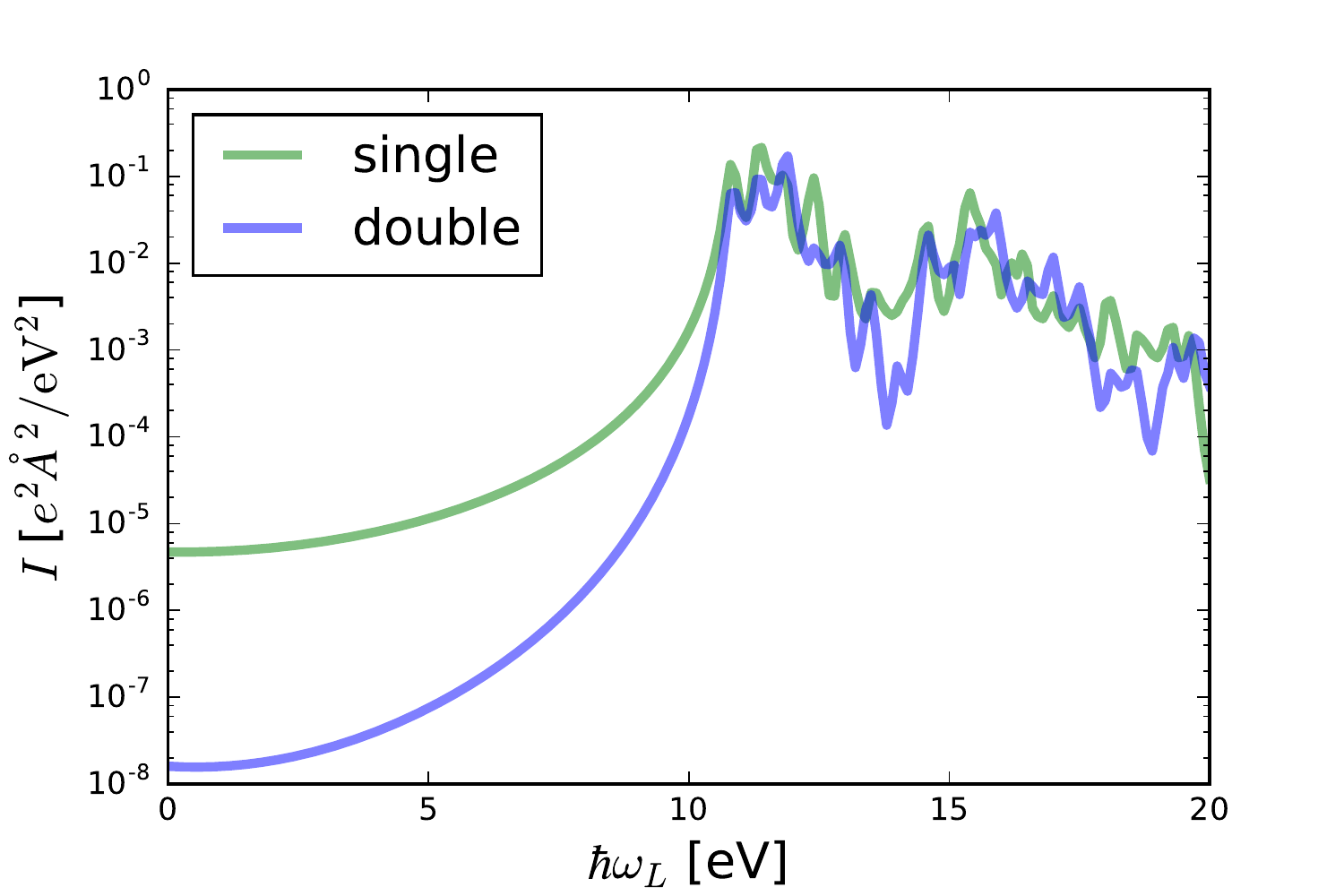}}
  \caption{Single and double vibrational excitation intensities
  according to the Albrecht A term in the H$_2$ molecule.
  A width of $\gamma=0.1$ eV is applied.
  }
  \label{fig:H2multiples}
\end{figure}
Now we turn to multiple vibrational excitations\ch{, 
the so called overtones and combinations bands in contrast to 
the fundamentals that correspond to single excitations. \cite{Neese07ccr}} 
Here, we expect 
severe difference between Albrecht and Placzek
approximations since they are impossible in Placzek 
\ch{within the IMDHO approximation}.
As an example, the intensity of 
double vibrational excitation in the Albrecht A term 
of H$_2$ is compared to the single excitation Albrecht A term in
Fig. \ref{fig:H2multiples}. For low excitation energies
and thus far from resonance,
the intensity of vibrational double excitation is several orders of 
magnitude smaller than that of a single excitation.
This changes drastically near to resonances, where the intensities
become the same size and the intensity of the vibrational 
double excitation can even overbalance that of the single 
excitation. 

\begin{table}
  \begin{tabular}{l|c|c|c|c|c|c|c}
    & \multicolumn{2}{c|}{$\omega_v$} & \multicolumn{2}{c|}{$I_{\rm Ram}(0)$} &
    \multicolumn{3}{c}{$I_{\rm Ram}(\omega_L^{\rm exp}\equiv 514.5 {\rm nm})$}\\
    mode & ours & exp. & ours & others & ours & others & exp.$^f$ \\
    \hline
    $v_2(1A_1)$ & 1585 & 1595$^a$, 1638$^b$ & 1.4 & 0.8$^c$, 0.9$^d$, 1.1$^e$ &  1.4 & 1.1$^e$ & 0.9$\pm$0.2\\
    $v_1(2A_1)$ & 3747 & 3657$^a$, 3832$^b$ & 112 & 109$^c$, 120$^d$, 111$^e$ & 127 & 129$^e$ & 108$\pm$14\\
    $v_3(1B_2)$ & 3846 & 3756$^a$, 3943$^b$ & 25 & 26$^c$, 30$^d$, 26$^e$ & 28 & 29$^e$ & 19.2$\pm$2.1\\
    \hline
  \end{tabular}
  \\
  $^a$exp. vapor\cite{Benedict56jcp},
  $^b$exp. harmonic\cite{Johnson93jcp},
  $^c$PW92\cite{Porezag96prb},
  $^d$BP86\cite{Rappoport07},
  $^e$PBE\cite{Caillie00pccp},
  $^f$from ref. \cite{Stirling96jcp}
  \caption{\label{tab:H2OPlaczek}
    Vibrational frequencies in cm$^{-1}$ and
    Placzek absolute Raman intensities for \ch{$\hbar\omega_L/E_e\to 0$}
    in \AA$^4/$amu. A width of $\gamma=0.2$ eV is applied for the 
    calculations at $\omega_L^{\rm exp}$.
    }
\end{table}
Next we consider the water molecule.
We first discuss the results in the limit $\hbar\omega_L/E_e\to 0$
where several other calculations and extensive experimental data
(for small $\hbar\omega_L$)
are available.
The gas-phase water molecule has three independent
vibrational modes that are all Raman active.
Table \ref{tab:H2OPlaczek} shows the good agreement of our
calculated absolute Raman intensities in this limit
both with experiment as well as with other calculations.
There are differences due to different density functionals applied, but 
all approaches are of roughly the same good
accuracy as compared to experiment.
Static and dynamic polarizabilities for the experimental wavelength of
514.5 nm (2.41 eV) lead to small differences \cite{Caillie00pccp}, only.
While the very weak Raman intensity of $v_2$ does not change, the stronger
$v_1$ and $v_3$ slightly increase.
\begin{table}
  \begin{tabular}{c|r|r|r|r|r|r|r}
    mode & exp\cite{Stirling96jcp} & Albrecht & Albrecht \cite{Gong15jctc}&
    Albr. $A$ & Albr. $A$ \cite{Gong15jctc} &
    $B+C$ & $B+C$ \cite{Gong15jctc} \\
    \hline
    $v_2$ & 0.9  &  2.7 &  1.1 & 0.5 &  0.1 & 5.3 & 1.5\\
    $v_1$ & 108  & 103 & 105 & 7.6 & 11.7 & 56 & 48.3\\
    $v_3$ & 19.2 &  30 & 25.7 &   0 &    0 & 30 & 25.7\\
    \hline
  \end{tabular}
  \\
  \caption{\label{tab:H2OAlbrecht}
    Absolute Raman intensities for $\omega_L=0$ in the Albrecht approximations
    from our calculations
    and from Gong et al.\cite{Gong15jctc}
    in \AA$^4/$amu.
    }
\end{table}
Table \ref{tab:H2OAlbrecht} lists the static 
($\omega_L\to 0$) absolute Raman contributions
from the Albrecht terms. Our calculation agrees well with the
the recent results of Gong et al.\cite{Gong15jctc} and we 
confirm that the Albrecht A term cannot be 
neglected in this limit for $v_1$ and $v_2$.

\begin{figure}[htbp]
  \centering{
    \includegraphics[width=0.7\textwidth]{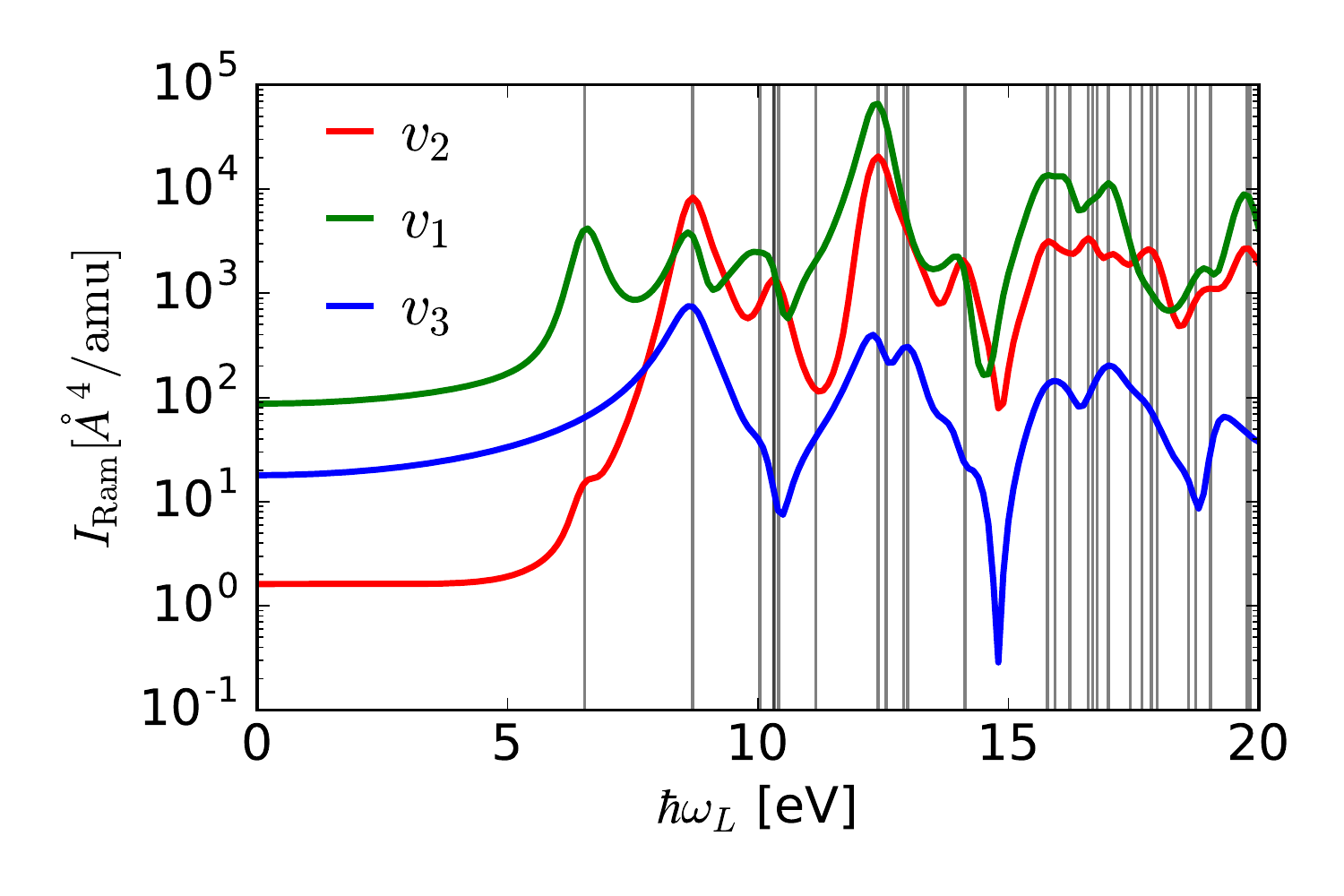}}
  \caption{
    Frequency dependent Placzek-Raman intensities of the three vibrations 
    in water. A width of $\gamma=0.2$ eV is applied.
    Optically active transitions with $f \ge 0.02$ are marked by gray lines.
  }
  \label{H2Oraman}
\end{figure}
Next we increase the energy of the incoming photon 
to the resonance
region as shown in Fig. \ref{H2Oraman},
where $I_{\rm Ram}$ is depicted on logarithmic 
scale in the region up to 20 eV. 
There are dramatic changes 
in the relative intensities.
Coming from small $\omega_L$ all three vibrations show increasing intensity
when entering the resonance region.  
The behavior of the three vibrations is very different, however. While
$v_1$ and $v_2$ show a clear peak at the first optically active 
transition, $v_3$ is unaffected.
Interestingly, vibration $v_1$, that is extremely weak in the limit 
$\omega_L\to 0$
increases most rapidly and even gets the highest contribution around 9 eV.
The deep minima in the intensities visible in particular
for $v_3$ indicate destructive interference that strongly affects
the intensity.

Finally we present the non-resonant and near-resonant Raman 
spectra of trans-butadiene where experimental non-resonant\cite{Richards50}
resonant\cite{Chadwick91jcp} Raman spectra 
as well as an early theoretical investigation within the Albrecht 
approximation\cite{Warshel77jcp} are available.
\ch{Duschinsky effects have been reported in this molecule \cite{Phillips93jpc}
  and these are important to understand the 
  fluorescence quantum efficiency.\cite{Krawczyk00cpl, Peng07jacs}
  The neglect of mode mixing as in the IMDHO model applied here has been
  reported to capture the geometry changes in the main 
  optical $1 ^1A_g^-\to ^1B_u^+$ transition,\cite{Hemley83jcp} however.
  }
In the experiment this 
transition \ch{is found} 
at 5.92 eV. Our calculated value (5.53 eV) is lower than 
the experimental one as well as the transition energy in higher level quantum chemistry methods\cite{Hsu01jpca}. 
The oscillator strength is 0.68, \ch{which is} higher
than the experimental value of 0.4\cite{Hsu01jcpa}, but in concord with other
computations\cite{Hsu01jcpa}. The underestimation of excitation energies
in GGAs is well known\cite{Grimme04}, while the rather accurate oscillator 
strength indices a qualitatively correct description of the
nature of the excitation. 
In order to compare our resonant Raman calculations
to experiment we have to take account of the differences
in excitation energies. 
Multiple vibrational excitations appear only if there is
a nearby resonance and their contribution
crucially depends on the energetic distance between $\hbar\omega_L$
and $E_I$ as we have seen in Fig. \ref{fig:H2multiples} for H$_2$.
In the following we  calculate our spectra
at 5.92 eV - 5.53 eV = 0.39 eV lower excitation energies than
in experiment in order to be at the same energetic 
distance from the main optical resonance.

\begin{figure}[h]
  \centering{
    \includegraphics[width=\textwidth]{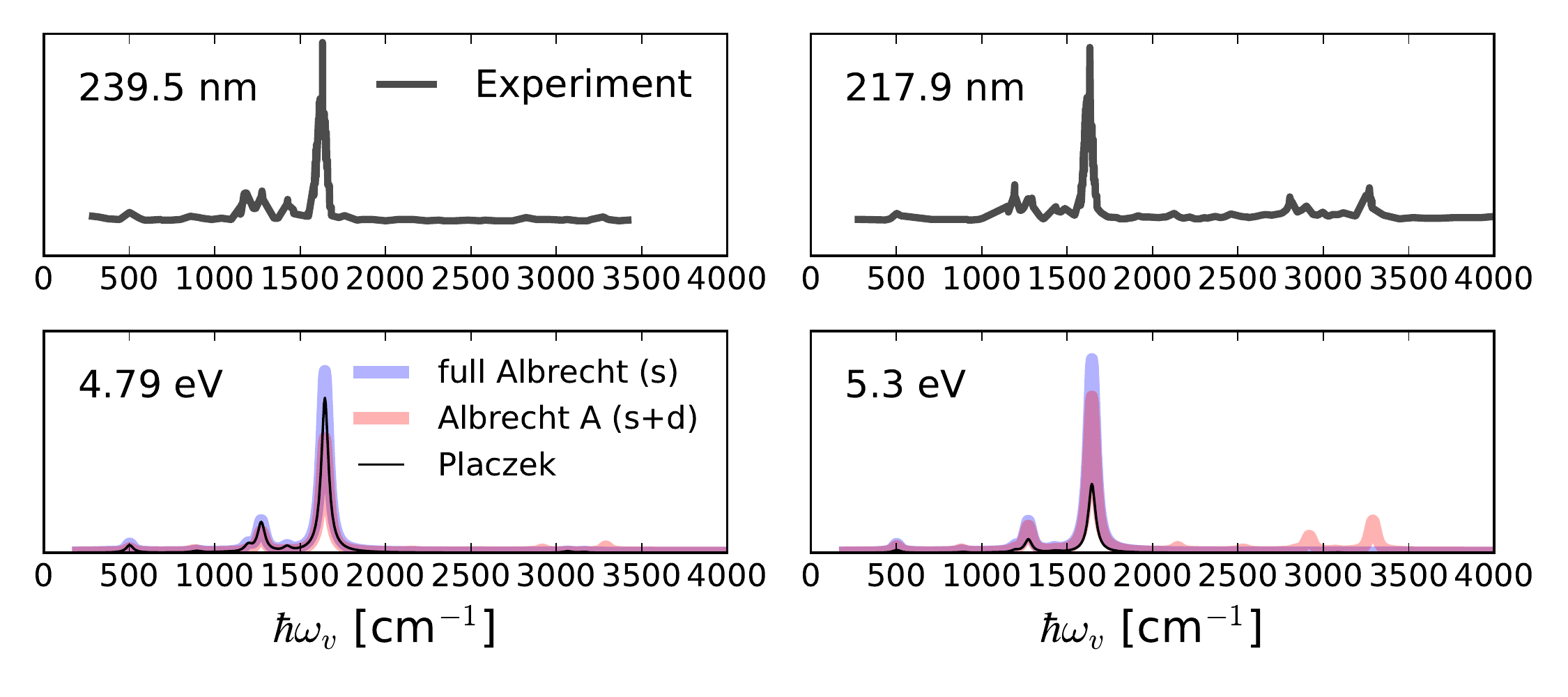}}
  \caption{Near resonant ($\hbar\omega_L=5.3$ eV 
    $\mathrel{\widehat{=}}$ 217.9 nm, see text) and more off-resonant 
    ($\hbar\omega_L=4.79$ eV $\mathrel{\widehat{=}}$ 239.5 nm) 
    Raman spectra for trans-butadiene
    in comparison to the experimental spectra of Chadwick et 
    al.\cite{Chadwick91jcp}
    The spectrum labeled full Albrecht (s) contains all Albrecht A-C terms for 
    single vibrational contributions, while the  
    Albrecht A (s+d) term contains vibrational double excitations also.
    A width of $\gamma=0.2(0.4)$ eV was applied to Albrecht(Placzek) 
    approximations and the intensities were folded
    by Lorentzians of 50 cm$^{-1}$ width.
  }
  \label{fig:Butadiene}
\end{figure}
Experimental and calculated spectra related in this way closely resemble
each other as is shown in Fig. \ref{fig:Butadiene}.
The spectrum in the non-resonant region (at 4.79 eV in our calculation
corresponding to 239.5 nm or 5.18 eV
experimentally)  has significant contribution from 
the Albrecht B and C terms  and there is nearly no intensity
above 2000 cm$^{-1}$. 
In this frequency range the Placzek approximation gives a 
good description of 
the spectrum. \ch{Apart from the main peak at 1645 cm$^{-1}$ 
also other small peaks at lower vibrational energies appear 
in the calculation similar to experiment
as further detailed in Tab.~\ref{tab:Butadiene} below.}
This changes near to resonance (5.30 eV in our calculation
corresponding to 
217.9 nm or 5.69 eV experimentally), 
where Albrecht A practically determines the full intensity.
The Placzek approximation gives much lower intensity as compared to 
Albrecht. 
There are also contributions of higher excitations due to the dominating
Albrecht A term
that naturally appear at higher vibrational frequencies.
These are completely neglected within the Placzek approximation.

\begin{table}
  \begin{tabular}{c|r|l|r|r|r|r}
    $\hbar\omega_v$ & $\hbar\omega_v$ (exp) & assignment 
    & $I^{\rm Albrecht}$ & $I^{\rm Placzek}$ & $I^{\rm Albrecht A}$ & $I^{\rm Albrecht, c}$\\
    \hline
    502  & 513$^a$, 508$^b$  & $v_9$ $\beta$(CCC)       & 4.1  & 3.8 & 3.9 & 0.2\\
    879  &  910$^a$, 890$^b$ & $v_8$ $\gamma$(C-C), $\beta$(H) & 0.8  & 1.1 & 1.9 & 1.3 \\
    1195 & 1204$^a$, 1200$^b$ & $v_7$ $\gamma$(C-C)    & 2.3  & 3.0 & 1.4 & 40.3 \\
    1272 & 1279$^a$, 1279$^b$ & $v_6$ $\beta$(H)       & 14.1 & 16.9 & 15 & 0.7 \\
    1275 &  & $\beta$(H)                       & 1.4 & 1.8 & 1.4 & -\\
    1424 & 1442$^a$, 1442$^b$ & $v_5$ $\beta$(H)       & 1.3 & 1.7 & 0.7 & 0.3 \\
    1645 & 1643$^a$, 1642$^b$ & $v_4$ $\gamma$(C-C,C=C) & 100(0.87) & 100(0.31) & 100(0.7) & 100 \\
    2917 & 2879$^a$ & $v_4+v_6$ & - & - & 10.7 & - \\
    3290 & 3267$^a$ & 2$v_4$ & - & - & 23.2 & - \\
    \hline
  \end{tabular}\\
   $^a$experiment of Richards and Nielsen\cite{Richards50}, $^b$experiment of 
   Chadwick et al.\cite{Chadwick91jcp}, $^c$ non-resonant 
   ($\hbar\omega_L=$1.24 eV) 
   Albrecht from Warshel and 
   Dauber\cite{Warshel77jcp}
   \caption{
  \label{tab:Butadiene}
  Vibrational energies (in cm$^-1$) and relative
  intensities of the most active Raman lines. 
  The intensity of the strongest line in $e^4\rm{\AA}^4/$eV$^2$
  is given in brackets.
  Assignment of modes follows Chadwick et al.\cite{Chadwick91jcp}, and
  $\beta, \gamma$ represent in-plane bending and stretching modes, 
  respectively.  
  Numerical settings as in Fig. \ref{fig:Butadiene}.
    }
\end{table}
The Raman peaks and their intensities are further detailed in 
tab. \ref{tab:Butadiene}. As suggested from Fig. \ref{fig:Butadiene} our
vibrational energies are in good agreement to experiment.
There are two in-plane H-bending modes at 1272 cm$^{-1}$ and 
1275 cm$^{-1}$ that are probably hard to resolve in the experiment.
One of them is the second most intense line in our calculations, while
the strong Raman excitation of the 1195 cm$^{-1}$ found in
the calculations of Warshel and Dauber\cite{Warshel77jcp}
cannot be confirmed by us (further off-resonant spectra resemble
the 4.79 eV spectrum in Fig. \ref{fig:Butadiene}).
Interestingly, the Placzek approximation gives relative intensities in 
good agreement to Albrecht despite its much lower 
absolute intensity.

\section{Conclusions}

We have shown in this contribution how the Placzek and Albrecht
terms can be derived from
the photon scattering matrix elements in the
formulation by Kramers, Heisenberg and Dirac.
The widely used Placzek approximation is found to
represent the semi-classical limit of the more exact 
Albrecht formulation. While the excitation energy-dependent
peak structure is much
simpler in the Placzek approximation, the overall
behavior of the Raman intensities is remarkably 
similar to Albrecht also in resonance regions \ch{for the molecules investigated}.
The similarity breaks down for multiple excitations
of vibrational modes appearing as soon as the 
photon energy approaches the resonance region.
These are forbidden in the Placzek form \ch{within the
  independent mode double harmonic approximation}, but are well described
within the Albrecht approximation.

\section{Acknowledgment}
  MW thanks J. Guthmuller for useful discussion.
  Support by ASTC is gratefully acknowledged. We are grateful for computational
  resources from FZ-J\"{u}lich and from the Nemo cluster at the University of
  Freiburg. 

\begin{appendix}
  \section{Connection between Placzek 
    and Albrecht approximations for $\omega_L\to 0$}
  \label{app:AlbrechtPlaczek}
  
  We consider the limit of very small excitation energies
  $\hbar\omega_L$,
  where small means far from any electronic resonance in the system.
  This limit can formally be described by $\omega_L\to 0$. 
  However, this is a formal limit only, as
  scattering of zero frequency photons is meaningless.
  Both Placzek and Albrecht approximations should be
  valid in this limit and the approximations indeed
  coincide as we will show in the following.

  The polarizability tensor (\ref{eq:alpha})
  in this limit simplifies to \cite{Hemert81mp}
  \begin{equation}
    \alpha(\omega = 0) = \sum_e \frac{2 m_L^e m_S^e}
          {E_{{\rm vert}, e}} \; .
  \end{equation}
  This tensor enters the Placzek approximation via
  its derivatives with respect to vibrational coordinates.
  Nonzero derivatives arise
  from two distinct sources: Either
  from the transition energy $E_{{\rm vert}, e}$ or from the
  matrix elements $m_{L,S}^e$.
  Without loss of generality, we consider only a single
  electronic excited state
  and thus suppress
  the label $e$ for brevity in the following.
  The explicit derivative is then
  \begin{equation}
    \frac{\partial\alpha}{\partial Q} =
    \frac{2}{E_{\rm vert}}\left[m_Sm_L' + m_Lm_S'\right] -
    2m_Sm_L\frac{E_{\rm vert}'}{E_{\rm vert}^2} \;,
    \label{eq:alphaQ0}
  \end{equation}
  where the prime denotes the derivative with respect to a nuclear coordinate.

  In order to show the equivalence to the Albrecht approximation
  for $\omega_L\to 0$, we will discuss the different terms
  appearing in (\ref{eq:alphaQ0}) separately and
  show that these correspond to the different terms in the
  Albrecht approximation, i.e. we will show that in the limit of 
  small $\omega_L$
  \begin{equation}
    V^A \approx  - \bra{0^0}Q\ket{1^0} 2m_Sm_L\frac{E_{\rm vert}'}{E_{\rm vert}^2} 
    \label{eq:VAeq}
  \end{equation}
  and
  \begin{equation}
    V^B+V^C \approx  \bra{0^0}Q\ket{1^0} \frac{2}{E_{\rm vert}}\left[m_Sm_L' + m_Lm_S'\right]
    \label{eq:VBCeq}
  \end{equation}
  hold. 

  We first discuss the Albrecht B term, 
  defined by eq. (\ref{eq:AlbrechtB}), which
  becomes in the limit of $\hbar\omega_L/E\to 0$ 
  \begin{equation}
    V^B = m_Sm_L'\sum_k\frac{\bra{0^0}Q\ket{k}\braket{k}{1^0}}{E+k\varepsilon}
    + m_Sm_L'\sum_k\frac{\braket{0^0}{k}\bra{k}Q\ket{1^0}}{E+k\varepsilon}\;
  \end{equation}
  Replacing the denominator $E + k\varepsilon$ by the vertical transition energy
  $E_{\rm vert}$ and applying the completeness relation
  eq. (\ref{eq:completeness}) leads to
  \begin{equation}
    V^B = \frac{1}{E_{\rm vert}}\left[m_Sm_L' + m_Lm_S'\right]\bra{0^0}Q\ket{1^0}
  \end{equation}
  which is the half of the first term in eq. (\ref{eq:VBCeq}). The
  other half is given by $V^C$ within the same approximation.

  To prove (\ref{eq:VAeq}) we consider real matrix elements
  \begin{equation}
    V^A=m_Lm_S\sum_k \braket{0^0}{k}\braket{k}{1^0}
    \left[\frac{1}{E+k\varepsilon-\omega_L} +
      \frac{1}{E+k\varepsilon+\omega_S}\right]\; ,
  \end{equation}
  use $\omega_S\approx\omega_L$, $\omega_L\to 0$ and expand
  in $\epsilon/E$ up to first order, which leads to
  \begin{equation}
    V^A=2 m_Lm_S\sum_k \braket{0^0}{k}\braket{k}{1^0}
    \frac{1}{E}\left(1 - k
      \frac{\varepsilon}{E}\right)
    \label{eq:VA0}
  \end{equation}
  The first term in brackets vanishes due to orthogonality of
  $\braket{0^0}{1^0}$ (after closure in $k^e$) and one can show
  that
  \begin{equation}
    \sum_k \braket{0^0}{k}k\braket{k}{1^0} = 
    -\frac{\Delta}{\sqrt{2}}
  \end{equation}
  where the dimensionless displacement
  \begin{equation}
    \Delta=\sqrt{\frac{\mu\omega}{\hbar}}\left(\xi_0^{(e)}-\xi_0^{(0)}\right)
  \end{equation}
  appears (c.f. Fig. \ref{fig:DisplacedHO}). One can further show that 
  \begin{equation}
    \bra{0^0}Q\ket{1^0}E_{\rm vert}'=\sqrt{\frac{\hbar}{2\omega}}
    \left(-\varepsilon \Delta \sqrt{\frac{\omega}{\hbar}}\right)
  \end{equation}
  such that eq. (\ref{eq:VA0}) can be written
  \begin{equation}
    V^A \approx -2 m_Lm_S \frac{E_{\rm vert}'}{E_{\rm vert}^2}
  \end{equation}
  where $E\approx E_{\rm vert}$ entered. 
  This finally proofs the approximate 
  equality of Placzek and Albrecht approximations in the
  limit $\omega_L\to 0$.

  \section{Taylor expansion and displaced harmonic 
    oscillator}
  \label{appFC}

  Franck-Condon factors can be efficiently calculated within the double
  harmonic approximation. 
  Deviations from this approximation, the Herzberg-Teller and Duschinsky 
  effects are usually rather small\cite{Guo12}.
  This property can be understood by expanding the possible 
  effects in a series in the displacement between ground and 
  excited state equilibria $\rho_0$ and $\rho_e$ of some vibrational 
  coordinate $\rho$. 
  The ground state potential in the harmonic approximation is given by
  \begin{eqnarray}
    E_0(\rho) = E_0(\rho_0) + \frac{1}{2}\mu\omega^2(\rho-\rho_0)^2 \; ,
    \label{eq:Eo}
  \end{eqnarray}
  where $\mu$ is the effective mass and $\omega$ the 
  corresponding frequency of the 
  harmonic potential. 
  We may expand the excited state energy around $\rho_0$
  in a Taylor series up to first order
  \begin{eqnarray}
    E_e(\rho) = E_e(\rho_0) + 
  \left.\frac{\partial E_e(\rho)}{\partial\rho}\right|_{\rho=\rho_0} (\rho-\rho_0)
  + O[(\rho-\rho_0)^2]
  \label{eq:Ee}
  \end{eqnarray}
  Adding Eqs. (\ref{eq:Eo}) and (\ref{eq:Ee}) immediately leads to the 
  similar harmonic equation in the excited state
  \begin{eqnarray}
    E_e(\rho) = E_e(\rho_0) + \frac{1}{2}\mu\omega^2(\rho-\rho_e)^2
    \label{eq:Eoex}
  \end{eqnarray}
  where we have identified 
  $\rho_e=\rho_0-\frac{\partial E_e(\rho)}{\partial\rho}/(\mu\omega^2)$.
  Thus the leading term in eq. (\ref{eq:Ee}) changes 
  the equilibrium position, but not the vibrational frequency\cite{Keil65}.

  \section{Matrix element derivatives and 
    the Berry phase}
  \label{app:Berry}
  
  The evaluation of Albrecht B and C terms requires
  derivatives of transition dipoles with respect to nuclear coordinates.
  An evaluation of such derivatives in finite differences is not
  straightforward as it involves arbitrary phases that are present
  in eigenstates of parameter dependent Hamiltonians and are
  connected to the Berry phase \cite{Resta00jpc,Min14prl}.
  Similar problems are also present in the evaluation of
  hopping matrix elements \cite{Baer02jcp}.

  The nature of the problem and its solution can be exemplified 
  in one dimension involving
  a single electronic positional coordinate $x$.
  An electron might be subject to a 
  parameter dependent Hamiltonian $H(R)$, where in
  $R$ in our case is a nuclear coordinate. The aim is to calculate
  the derivative of an normalized eigenstate
  $f_i(x;R)$ of $H(R)$ with respect
  to the $R$ in a finite difference expression
  \begin{equation}
     dR \frac{\partial f_i(x;R)}{\partial R} = 
     f_i(x; R+dR) - f_i(x;R) \; .
     \label{eq:fd}
  \end{equation}
  In practical calculations the evaluation
  of eigenstates at different $R$ are independent of each 
  other\cite{Baer02jcp}. Then
  every eigenstate $\bar{f}_i(x;R+dR) = uf_i(x;R+dR)$ with $|u|^2=1$ is
  a perfectly valid eigenstate of $H(R+dR)$ and equally 
  relevant as $f_i(x;R+dR)$ itself. The phase $u$ can
  spoil the derivative if $\bar{f}$ is used
  instead of $f$ in expression (\ref{eq:fd}), however.
  To recover $f_i(x;R+dR)$ we have to apply 
  \begin{equation}
     dR \frac{\partial f_i(x;R)}{\partial R} = 
     u^*\bar{f}_i(x; R+dR) - f_i(x;R)
     \label{eq:fdph}
  \end{equation}
  instead, where we correct for the arbitrary phase $u$. The value of this 
  phase factor is reconstructed by using the orthogonality
  of $f_i$ and $\partial f_i / \partial R$ \cite{Resta00jpc}
  that is required for normalized states. It leads to
  \begin{equation}
    u = \int dx\, f^*(x;R)\bar{f}(x;R+dR) \; .
    \label{eq:uval}
  \end{equation}

  We have to be slightly more careful in the case of 
  energetically degenerate states, that are common in molecules. 
  Here, not only a phase $u$ may appear, but the states may also mix.
  More generally we are faced with
  \begin{equation}
    \bar{f}_i(x;R+dR) = \sum_j u_{ij} f_j(x;R+dR)
  \end{equation}
  where the matrix $u$ is unitary, but arbitrary otherwise.
  It might be sparse, but generally not diagonal.
  Similar to (\ref{eq:uval}) the matrix elements of 
  $u$ can be reconstructed from
  \begin{equation}
    u_{ij} = \int dx\, f_j^*(x;R)\bar{f}_i(x;R+dR)
    \label{eq:uij}
  \end{equation}
  when terms containing $dR$ are neglected, i.e. we assume that
  the matrix $u$ does not change due to the small displacement.
  This leads to the generalized finite difference equation
  save from arbitrary phases
  \begin{equation}
     dR \frac{\partial f(x;R)}{\partial R} = 
     u^H\bar{f}(x; R+dR) - f(x;R) \; .
     \label{eq:gfd}
  \end{equation}
  where $f$ is the vector of eigenstates $f_i$, 
  $\bar{f}$ is the vector of eigenstates $\bar{f}_j$
  and the superscript $H$ denotes the Hermitian conjugate.

  Similar to the Eigenstates discussed so far, we want to
  obtain derivatives
  of transition dipole
  matrix elements $m_{i\alpha}(R)=\bra{f_i(x;R)}\hat{o}\ket{f_\alpha(x;R)}$
  in a finite difference expression through
  \begin{equation}
    dR \frac{\partial m_{i\alpha}(R)}{\partial R} = m_{i\alpha}(R+dR) -
    m_{i\alpha}(R) \; ,
  \end{equation}
  where $i, \alpha$ are the indices of occupied and empty orbitals, 
  respectively. The transition dipoles are 
  calculated in an independent calculation again and thus are 
  mixed and contain arbitrary phases inherited from the orbitals.
  We may correct for this similar to eq. (\ref{eq:gfd}) and write
  \begin{equation}
    dR \frac{\partial m_{i\alpha}(R)}{\partial R} =
    \bra{\left[u^{H}\bar{f}(R+dR)\right]_i}\hat{o}
    \ket{\left[u^H\bar{f}(R+dR)\right]_\alpha} -
    m_{i\alpha}(R)
    \label{eq:mia}
  \end{equation}
  or in vector form
  \begin{equation}
    dR \frac{\partial m(R)}{\partial R} = U^H \bar{m}(R+dR) - m(R)
    \label{eq:dmdRKS}
  \end{equation}
  with
  \begin{equation}
    U_{i\alpha, j\beta} = u_{ij}^*u_{\alpha\beta}
    \label{eq:Upq}
  \end{equation}
  and $\bar{m}_{i\alpha}(R+dR)=\bra{\bar{f}_i(x;R+dR)}\hat{o}\ket{\bar{f}_\alpha(x;R+dR)}$.

  A new class of phases appears in linear response TDDFT where the
  eigenvalue equation\cite{Casida09,Walter08jcp}
  \begin{equation}
    \Omega F_I = \omega_I^2 F_I   
  \end{equation}
  is solved at each position independently. 
  The $\omega_I$ denote
  transition energies and the eigenvectors $F_I$ may contain 
  arbitrary phases and might be mixed. 
  The matrix elements are then
  \begin{equation}
     M_I=\sum_{i\alpha}\sqrt{
     \frac{\varepsilon_\alpha-\varepsilon_i}{\omega_I}}
     (F_I)_{i\alpha} m_{i\alpha} 
  \end{equation}
  with the single particle energies $\varepsilon_{i,\alpha}$.
  The $F_I$ at equilibrium 
  position and the $\bar{F}_J$ at a displaced position 
  are given in the corresponding particle-hole basis
  that we may contract to single indices
  $p=(i\alpha), q=(j,\beta)$ to simplify the notation.
  We may define an overlap similar to Eqs. (\ref{eq:uij})
  and (\ref{eq:Upq})
  \begin{equation}
    W_{IJ} = \sum_{p, q} (\bar{F}_{J})_pU_{pq}(F_{I})_q^*
    \label{eq:WIJ}
  \end{equation}
  where the $U_{pq}$ are needed to connected the two particle-hole
  bases. 
  This matrix connects the different linear response
  transition matrix elements
  via $\bar{M}=W M$, where
  \begin{equation}
    \bar{M}_I=\sum_{i\alpha}\sqrt{
     \frac{\varepsilon_\alpha-\varepsilon_i}{\omega_I}}
     (\bar{F}_I)_{i\alpha} \bar{m}_{i\alpha} \; .
  \end{equation}
  Note, that the phases of $\bar{F}$ and $\bar{m}$ are both arbitrary
  and independent of each other.
  Derivatives of linear response dipole matrix elements are 
  finally obtained as
  \begin{equation}
    dR \frac{\partial M(R)}{\partial R} = W^H \bar{M}(R+dR) - m(R) \; .
    \label{eq:dmdRLR}
  \end{equation}

\end{appendix}

\bibliography{rraman}

\providecommand{\latin}[1]{#1}
\makeatletter
\providecommand{\doi}
  {\begingroup\let\do\@makeother\dospecials
  \catcode`\{=1 \catcode`\}=2 \doi@aux}
\providecommand{\doi@aux}[1]{\endgroup\texttt{#1}}
\makeatother
\providecommand*\mcitethebibliography{\thebibliography}
\csname @ifundefined\endcsname{endmcitethebibliography}
  {\let\endmcitethebibliography\endthebibliography}{}
\begin{mcitethebibliography}{90}
\providecommand*\natexlab[1]{#1}
\providecommand*\mciteSetBstSublistMode[1]{}
\providecommand*\mciteSetBstMaxWidthForm[2]{}
\providecommand*\mciteBstWouldAddEndPuncttrue
  {\def\EndOfBibitem{\unskip.}}
\providecommand*\mciteBstWouldAddEndPunctfalse
  {\let\EndOfBibitem\relax}
\providecommand*\mciteSetBstMidEndSepPunct[3]{}
\providecommand*\mciteSetBstSublistLabelBeginEnd[3]{}
\providecommand*\EndOfBibitem{}
\mciteSetBstSublistMode{f}
\mciteSetBstMaxWidthForm{subitem}{(\alph{mcitesubitemcount})}
\mciteSetBstSublistLabelBeginEnd
  {\mcitemaxwidthsubitemform\space}
  {\relax}
  {\relax}

\bibitem[{Derek A. Long}(2002)]{Long02}
{Derek A. Long}, \emph{The {Raman} {Effect}: {A} {Unified} {Treatment} of the
  {Theory} of {Raman} {Scattering} by {Molecules}}; John Wiley \& Sons Ltd,
  Baffins Lane, Chichester, West Sussex PO19 1UD, England, 2002\relax
\mciteBstWouldAddEndPuncttrue
\mciteSetBstMidEndSepPunct{\mcitedefaultmidpunct}
{\mcitedefaultendpunct}{\mcitedefaultseppunct}\relax
\EndOfBibitem
\bibitem[{John R. Ferraro} \latin{et~al.}(2003){John R. Ferraro}, {Kazuo
  Nakamoto}, and {Chris W. Brown}]{Ferraro03}
{John R. Ferraro},; {Kazuo Nakamoto},; {Chris W. Brown}, \emph{Introductory
  {Raman} {Spectroscopy} ({Second} {Edition})}; Elsevier Inc., 2003\relax
\mciteBstWouldAddEndPuncttrue
\mciteSetBstMidEndSepPunct{\mcitedefaultmidpunct}
{\mcitedefaultendpunct}{\mcitedefaultseppunct}\relax
\EndOfBibitem
\bibitem[Porezag and Pederson(1996)Porezag, and Pederson]{Porezag96prb}
Porezag,~D.; Pederson,~M.~R. Infrared intensities and {Raman}-scattering
  activities within density-functional theory. \emph{Physical Review B}
  \textbf{1996}, \emph{54}, 7830--7836\relax
\mciteBstWouldAddEndPuncttrue
\mciteSetBstMidEndSepPunct{\mcitedefaultmidpunct}
{\mcitedefaultendpunct}{\mcitedefaultseppunct}\relax
\EndOfBibitem
\bibitem[Yamakita \latin{et~al.}(2007)Yamakita, Kimura, and
  Ohno]{Yamakita07jcp}
Yamakita,~Y.; Kimura,~J.; Ohno,~K. Molecular vibrations of [n]oligoacenes
  (n=2-5 and 10) and phonon dispersion relations of polyacene. \emph{The
  Journal of Chemical Physics} \textbf{2007}, \emph{126}, 064904\relax
\mciteBstWouldAddEndPuncttrue
\mciteSetBstMidEndSepPunct{\mcitedefaultmidpunct}
{\mcitedefaultendpunct}{\mcitedefaultseppunct}\relax
\EndOfBibitem
\bibitem[Castiglioni \latin{et~al.}(2004)Castiglioni, Tommasini, and
  Zerbi]{Castiglioni04}
Castiglioni,~C.; Tommasini,~M.; Zerbi,~G. Raman spectroscopy of polyconjugated
  molecules and materials: confinement effect in one and two dimensions.
  \emph{Philosophical Transactions of the Royal Society of London A:
  Mathematical, Physical and Engineering Sciences} \textbf{2004}, \emph{362},
  2425--2459\relax
\mciteBstWouldAddEndPuncttrue
\mciteSetBstMidEndSepPunct{\mcitedefaultmidpunct}
{\mcitedefaultendpunct}{\mcitedefaultseppunct}\relax
\EndOfBibitem
\bibitem[Rouillé \latin{et~al.}(2008)Rouillé, Jäger, Steglich, Huisken,
  Henning, Theumer, Bauer, and Knölker]{Rouille08}
Rouillé,~G.; Jäger,~C.; Steglich,~M.; Huisken,~F.; Henning,~T.; Theumer,~G.;
  Bauer,~I.; Knölker,~H.-J. {IR}, {Raman}, and {UV}/{Vis} {Spectra} of
  {Corannulene} for {Use} in {Possible} {Interstellar} {Identification}.
  \emph{ChemPhysChem} \textbf{2008}, \emph{9}, 2085--2091\relax
\mciteBstWouldAddEndPuncttrue
\mciteSetBstMidEndSepPunct{\mcitedefaultmidpunct}
{\mcitedefaultendpunct}{\mcitedefaultseppunct}\relax
\EndOfBibitem
\bibitem[Zhao \latin{et~al.}(2006)Zhao, Jensen, and Schatz]{Zhao06nl}
Zhao,~L.~L.; Jensen,~L.; Schatz,~G.~C. Surface-{Enhanced} {Raman} {Scattering}
  of {Pyrazine} at the {Junction} between {Two} {Ag}20 {Nanoclusters}.
  \emph{Nano Letters} \textbf{2006}, \emph{6}, 1229--1234\relax
\mciteBstWouldAddEndPuncttrue
\mciteSetBstMidEndSepPunct{\mcitedefaultmidpunct}
{\mcitedefaultendpunct}{\mcitedefaultseppunct}\relax
\EndOfBibitem
\bibitem[Martin \latin{et~al.}(2015)Martin, Bérubé, Provencher, Côté,
  Silva, Doorn, and Grey]{Martin15jmcc}
Martin,~E. J.~J.; Bérubé,~N.; Provencher,~F.; Côté,~M.; Silva,~C.;
  Doorn,~S.~K.; Grey,~J.~K. Resonance {Raman} spectroscopy and imaging of
  push–pull conjugated polymer–fullerene blends. \emph{Journal of Materials
  Chemistry C} \textbf{2015}, \emph{3}, 6058--6066\relax
\mciteBstWouldAddEndPuncttrue
\mciteSetBstMidEndSepPunct{\mcitedefaultmidpunct}
{\mcitedefaultendpunct}{\mcitedefaultseppunct}\relax
\EndOfBibitem
\bibitem[Vecera \latin{et~al.}(2017)Vecera, Chacón-Torres, Pichler, Reich,
  Soni, Görling, Edelthalhammer, Peterlik, Hauke, and Hirsch]{Vecera17nat}
Vecera,~P.; Chacón-Torres,~J.~C.; Pichler,~T.; Reich,~S.; Soni,~H.~R.;
  Görling,~A.; Edelthalhammer,~K.; Peterlik,~H.; Hauke,~F.; Hirsch,~A. Precise
  determination of graphene functionalization by in situ {Raman} spectroscopy.
  \emph{Nature Communications} \textbf{2017}, \emph{8}, 15192\relax
\mciteBstWouldAddEndPuncttrue
\mciteSetBstMidEndSepPunct{\mcitedefaultmidpunct}
{\mcitedefaultendpunct}{\mcitedefaultseppunct}\relax
\EndOfBibitem
\bibitem[Corni \latin{et~al.}(2001)Corni, Cappelli, Cammi, and
  Tomasi]{Corni01jcpa}
Corni,~S.; Cappelli,~C.; Cammi,~R.; Tomasi,~J. Theoretical {Approach} to the
  {Calculation} of {Vibrational} {Raman} {Spectra} in {Solution} within the
  {Polarizable} {Continuum} {Model}. \emph{The Journal of Physical Chemistry A}
  \textbf{2001}, \emph{105}, 8310--8316\relax
\mciteBstWouldAddEndPuncttrue
\mciteSetBstMidEndSepPunct{\mcitedefaultmidpunct}
{\mcitedefaultendpunct}{\mcitedefaultseppunct}\relax
\EndOfBibitem
\bibitem[Cheeseman and Frisch(2011)Cheeseman, and Frisch]{Cheeseman11}
Cheeseman,~J.~R.; Frisch,~M.~J. Basis {Set} {Dependence} of {Vibrational}
  {Raman} and {Raman} {Optical} {Activity} {Intensities}. \emph{Journal of
  Chemical Theory and Computation} \textbf{2011}, \emph{7}, 3323--3334\relax
\mciteBstWouldAddEndPuncttrue
\mciteSetBstMidEndSepPunct{\mcitedefaultmidpunct}
{\mcitedefaultendpunct}{\mcitedefaultseppunct}\relax
\EndOfBibitem
\bibitem[Barone \latin{et~al.}(2014)Barone, Biczysko, and Bloino]{Barone14pccp}
Barone,~V.; Biczysko,~M.; Bloino,~J. Fully anharmonic {IR} and Raman spectra of
  medium-size molecular systems: accuracy and interpretation. \emph{Phys. Chem.
  Chem. Phys.} \textbf{2014}, \emph{16}, 1759--1787\relax
\mciteBstWouldAddEndPuncttrue
\mciteSetBstMidEndSepPunct{\mcitedefaultmidpunct}
{\mcitedefaultendpunct}{\mcitedefaultseppunct}\relax
\EndOfBibitem
\bibitem[Bloino \latin{et~al.}(2015)Bloino, Biczysko, and Barone]{Bloino15jpca}
Bloino,~J.; Biczysko,~M.; Barone,~V. Anharmonic {Effects} on {Vibrational}
  {Spectra} {Intensities}: {Infrared}, {Raman}, {Vibrational} {Circular}
  {Dichroism}, and {Raman} {Optical} {Activity}. \emph{The Journal of Physical
  Chemistry A} \textbf{2015}, \emph{119}, 11862--11874\relax
\mciteBstWouldAddEndPuncttrue
\mciteSetBstMidEndSepPunct{\mcitedefaultmidpunct}
{\mcitedefaultendpunct}{\mcitedefaultseppunct}\relax
\EndOfBibitem
\bibitem[Ambrosch-Draxl \latin{et~al.}(2002)Ambrosch-Draxl, Auer, Kouba,
  Sherman, Knoll, and Mayer]{Ambrosch02prb}
Ambrosch-Draxl,~C.; Auer,~H.; Kouba,~R.; Sherman,~E.~Y.; Knoll,~P.; Mayer,~M.
  Raman scattering in YBa$_2$Cu$_3$O$_7$: A comprehensive theoretical study in
  comparison with experiments. \emph{Physical Review B} \textbf{2002},
  \emph{65}, 064501\relax
\mciteBstWouldAddEndPuncttrue
\mciteSetBstMidEndSepPunct{\mcitedefaultmidpunct}
{\mcitedefaultendpunct}{\mcitedefaultseppunct}\relax
\EndOfBibitem
\bibitem[Gillet \latin{et~al.}(2013)Gillet, Giantomassi, and
  Gonze]{Gillet13prb}
Gillet,~Y.; Giantomassi,~M.; Gonze,~X. First-principles study of excitonic
  effects in {Raman} intensities. \emph{Physical Review B} \textbf{2013},
  \emph{88}, 094305\relax
\mciteBstWouldAddEndPuncttrue
\mciteSetBstMidEndSepPunct{\mcitedefaultmidpunct}
{\mcitedefaultendpunct}{\mcitedefaultseppunct}\relax
\EndOfBibitem
\bibitem[{Li Niu} \latin{et~al.}(2008){Li Niu}, {Jiaqi Zhu}, {Wei Gao}, {Aiping
  Liu}, {Xiao Han}, and {Shanyi Du}]{Niu08pb}
{Li Niu},; {Jiaqi Zhu},; {Wei Gao},; {Aiping Liu},; {Xiao Han},; {Shanyi Du},
  First-principles calculation of vibrational {Raman} spectra of tetrahedral
  amorphous carbon. \emph{Physica B: Condensed Matter} \textbf{2008},
  \emph{403}, 3559--3562\relax
\mciteBstWouldAddEndPuncttrue
\mciteSetBstMidEndSepPunct{\mcitedefaultmidpunct}
{\mcitedefaultendpunct}{\mcitedefaultseppunct}\relax
\EndOfBibitem
\bibitem[Wang \latin{et~al.}()Wang, Carvalho, and Crespi]{Wang18prb}
Wang,~Y.; Carvalho,~B.~R.; Crespi,~V.~H. Strong exciton regulation of Raman
  scattering in monolayer MoS$_2$. \emph{98}, 161405\relax
\mciteBstWouldAddEndPuncttrue
\mciteSetBstMidEndSepPunct{\mcitedefaultmidpunct}
{\mcitedefaultendpunct}{\mcitedefaultseppunct}\relax
\EndOfBibitem
\bibitem[Profeta and Mauri(2001)Profeta, and Mauri]{Profeta01prb}
Profeta,~M.; Mauri,~F. Theory of resonant {Raman} scattering of tetrahedral
  amorphous carbon. \emph{Physical Review B} \textbf{2001}, \emph{63},
  245415\relax
\mciteBstWouldAddEndPuncttrue
\mciteSetBstMidEndSepPunct{\mcitedefaultmidpunct}
{\mcitedefaultendpunct}{\mcitedefaultseppunct}\relax
\EndOfBibitem
\bibitem[Stock and Domcke(1990)Stock, and Domcke]{Stock90jcp}
Stock,~G.; Domcke,~W. Theory of resonance {Raman} scattering and fluorescence
  from strongly vibronically coupled excited states of polyatomic molecules.
  \emph{The Journal of Chemical Physics} \textbf{1990}, \emph{93},
  5496--5509\relax
\mciteBstWouldAddEndPuncttrue
\mciteSetBstMidEndSepPunct{\mcitedefaultmidpunct}
{\mcitedefaultendpunct}{\mcitedefaultseppunct}\relax
\EndOfBibitem
\bibitem[Peticolas and Rush(1995)Peticolas, and Rush]{Peticolas95jcc}
Peticolas,~W.~L.; Rush,~T. Ab initio calculations of the ultraviolet resonance
  {Raman} spectra of uracil. \emph{Journal of Computational Chemistry}
  \textbf{1995}, \emph{16}, 1261--1270\relax
\mciteBstWouldAddEndPuncttrue
\mciteSetBstMidEndSepPunct{\mcitedefaultmidpunct}
{\mcitedefaultendpunct}{\mcitedefaultseppunct}\relax
\EndOfBibitem
\bibitem[Jarzecki and Spiro(2001)Jarzecki, and Spiro]{Jarzecki01jrs}
Jarzecki,~A.~A.; Spiro,~T.~G. Ab initio computation of the {UV} resonance
  {Raman} intensity pattern of aqueous imidazole. \emph{Journal of Raman
  Spectroscopy} \textbf{2001}, \emph{32}, 599--605\relax
\mciteBstWouldAddEndPuncttrue
\mciteSetBstMidEndSepPunct{\mcitedefaultmidpunct}
{\mcitedefaultendpunct}{\mcitedefaultseppunct}\relax
\EndOfBibitem
\bibitem[Neugebauer and Hess(2004)Neugebauer, and Hess]{Neugebauer04jcp}
Neugebauer,~J.; Hess,~B.~A. Resonance {Raman} spectra of uracil based on
  {Kramers}–{Kronig} relations using time-dependent density functional
  calculations and multireference perturbation theory. \emph{The Journal of
  Chemical Physics} \textbf{2004}, \emph{120}, 11564--11577\relax
\mciteBstWouldAddEndPuncttrue
\mciteSetBstMidEndSepPunct{\mcitedefaultmidpunct}
{\mcitedefaultendpunct}{\mcitedefaultseppunct}\relax
\EndOfBibitem
\bibitem[Scholz \latin{et~al.}(2011)Scholz, Gisslén, Schuster, Casu, Chassé,
  Heinemeyer, and Schreiber]{Scholz11jcp}
Scholz,~R.; Gisslén,~L.; Schuster,~B.-E.; Casu,~M.~B.; Chassé,~T.;
  Heinemeyer,~U.; Schreiber,~F. Resonant {Raman} spectra of diindenoperylene
  thin films. \emph{The Journal of Chemical Physics} \textbf{2011}, \emph{134},
  014504\relax
\mciteBstWouldAddEndPuncttrue
\mciteSetBstMidEndSepPunct{\mcitedefaultmidpunct}
{\mcitedefaultendpunct}{\mcitedefaultseppunct}\relax
\EndOfBibitem
\bibitem[Balakrishnan \latin{et~al.}(2012)Balakrishnan, Jarzecki, Wu,
  Kozlowski, Wang, and Spiro]{Balakrishnan12jpcb}
Balakrishnan,~G.; Jarzecki,~A.~A.; Wu,~Q.; Kozlowski,~P.~M.; Wang,~D.;
  Spiro,~T.~G. Mode {Recognition} in {UV} {Resonance} {Raman} {Spectra} of
  {Imidazole}: {Histidine} {Monitoring} in {Proteins}. \emph{The Journal of
  Physical Chemistry B} \textbf{2012}, \emph{116}, 9387--9395\relax
\mciteBstWouldAddEndPuncttrue
\mciteSetBstMidEndSepPunct{\mcitedefaultmidpunct}
{\mcitedefaultendpunct}{\mcitedefaultseppunct}\relax
\EndOfBibitem
\bibitem[Warshel and Dauber(1977)Warshel, and Dauber]{Warshel77jcp}
Warshel,~A.; Dauber,~P. Calculations of resonance {Raman} spectra of conjugated
  molecules. \emph{The Journal of Chemical Physics} \textbf{1977}, \emph{66},
  5477--5488\relax
\mciteBstWouldAddEndPuncttrue
\mciteSetBstMidEndSepPunct{\mcitedefaultmidpunct}
{\mcitedefaultendpunct}{\mcitedefaultseppunct}\relax
\EndOfBibitem
\bibitem[Albrecht(1960)]{Albrecht60jcp}
Albrecht,~A.~C. ``{Forbidden}'' {Character} in {Allowed} {Electronic}
  {Transitions}. \emph{The Journal of Chemical Physics} \textbf{1960},
  \emph{33}, 156--169\relax
\mciteBstWouldAddEndPuncttrue
\mciteSetBstMidEndSepPunct{\mcitedefaultmidpunct}
{\mcitedefaultendpunct}{\mcitedefaultseppunct}\relax
\EndOfBibitem
\bibitem[Albrecht(1961)]{Albrecht61jcp}
Albrecht,~A.~C. On the {Theory} of {Raman} {Intensities}. \emph{The Journal of
  Chemical Physics} \textbf{1961}, \emph{34}, 1476--1484\relax
\mciteBstWouldAddEndPuncttrue
\mciteSetBstMidEndSepPunct{\mcitedefaultmidpunct}
{\mcitedefaultendpunct}{\mcitedefaultseppunct}\relax
\EndOfBibitem
\bibitem[Albrecht and Hutley(1971)Albrecht, and Hutley]{Albrecht71jcp}
Albrecht,~A.~C.; Hutley,~M.~C. On the {Dependence} of {Vibrational} {Raman}
  {Intensity} on the {Wavelength} of {Incident} {Light}. \emph{The Journal of
  Chemical Physics} \textbf{1971}, \emph{55}, 4438--4443\relax
\mciteBstWouldAddEndPuncttrue
\mciteSetBstMidEndSepPunct{\mcitedefaultmidpunct}
{\mcitedefaultendpunct}{\mcitedefaultseppunct}\relax
\EndOfBibitem
\bibitem[Myers~Kelley(2008)]{Myers08}
Myers~Kelley,~A. Resonance {Raman} and {Resonance} {Hyper}-{Raman}
  {Intensities}: {Structure} and {Dynamics} of {Molecular} {Excited} {States}
  in {Solution}. \emph{The Journal of Physical Chemistry A} \textbf{2008},
  \emph{112}, 11975--11991\relax
\mciteBstWouldAddEndPuncttrue
\mciteSetBstMidEndSepPunct{\mcitedefaultmidpunct}
{\mcitedefaultendpunct}{\mcitedefaultseppunct}\relax
\EndOfBibitem
\bibitem[Gong \latin{et~al.}(2015)Gong, Tian, Duan, and Luo]{Gong15jctc}
Gong,~Z.-Y.; Tian,~G.; Duan,~S.; Luo,~Y. Significant {Contributions} of the
  {Albrecht}’s {A} {Term} to {Nonresonant} {Raman} {Scattering} {Processes}.
  \emph{Journal of Chemical Theory and Computation} \textbf{2015}, \emph{11},
  5385--5390\relax
\mciteBstWouldAddEndPuncttrue
\mciteSetBstMidEndSepPunct{\mcitedefaultmidpunct}
{\mcitedefaultendpunct}{\mcitedefaultseppunct}\relax
\EndOfBibitem
\bibitem[Guthmuller(2016)]{Guthmuller16jcp}
Guthmuller,~J. Comparison of simplified sum-over-state expressions to calculate
  resonance {Raman} intensities including {Franck}-{Condon} and
  {Herzberg}-{Teller} effects. \emph{The Journal of Chemical Physics}
  \textbf{2016}, \emph{144}, 064106\relax
\mciteBstWouldAddEndPuncttrue
\mciteSetBstMidEndSepPunct{\mcitedefaultmidpunct}
{\mcitedefaultendpunct}{\mcitedefaultseppunct}\relax
\EndOfBibitem
\bibitem[{Eric J. Heller} \latin{et~al.}(2016){Eric J. Heller}, {Yuan Yang},
  {Lucas Kocia}, {Wei Chen}, {Shiang Fang}, {Mario Borunda}, and {Efthimios
  Kaxiras}]{Heller16an}
{Eric J. Heller},; {Yuan Yang},; {Lucas Kocia},; {Wei Chen},; {Shiang Fang},;
  {Mario Borunda},; {Efthimios Kaxiras}, Theory of {Graphene} {Raman}
  {Scattering}. \emph{ACS Nano} \textbf{2016}, \emph{10}, 2803--2818\relax
\mciteBstWouldAddEndPuncttrue
\mciteSetBstMidEndSepPunct{\mcitedefaultmidpunct}
{\mcitedefaultendpunct}{\mcitedefaultseppunct}\relax
\EndOfBibitem
\bibitem[Duan \latin{et~al.}(2016)Duan, Tian, and Luo]{Duan16jctc}
Duan,~S.; Tian,~G.; Luo,~Y. Theory for {Modeling} of {High} {Resolution}
  {Resonant} and {Nonresonant} {Raman} {Images}. \emph{Journal of Chemical
  Theory and Computation} \textbf{2016}, \emph{12}, 4986--4995\relax
\mciteBstWouldAddEndPuncttrue
\mciteSetBstMidEndSepPunct{\mcitedefaultmidpunct}
{\mcitedefaultendpunct}{\mcitedefaultseppunct}\relax
\EndOfBibitem
\bibitem[Hu \latin{et~al.}(2017)Hu, Duan, and Luo]{Hu16cms}
Hu,~W.; Duan,~S.; Luo,~Y. Theoretical modeling of surface and tip-enhanced
  {Raman} spectroscopies. \emph{WIREs Comput Mol Sci} \textbf{2017}, 1293\relax
\mciteBstWouldAddEndPuncttrue
\mciteSetBstMidEndSepPunct{\mcitedefaultmidpunct}
{\mcitedefaultendpunct}{\mcitedefaultseppunct}\relax
\EndOfBibitem
\bibitem[Avila~Ferrer \latin{et~al.}(2013)Avila~Ferrer, Barone, Cappelli, and
  Santoro]{Ferrer13jctc}
Avila~Ferrer,~F.~J.; Barone,~V.; Cappelli,~C.; Santoro,~F. Duschinsky,
  {Herzberg}–{Teller}, and {Multiple} {Electronic} {Resonance}
  {Interferential} {Effects} in {Resonance} {Raman} {Spectra} and {Excitation}
  {Profiles}. {The} {Case} of {Pyrene}. \emph{Journal of Chemical Theory and
  Computation} \textbf{2013}, \emph{9}, 3597--3611\relax
\mciteBstWouldAddEndPuncttrue
\mciteSetBstMidEndSepPunct{\mcitedefaultmidpunct}
{\mcitedefaultendpunct}{\mcitedefaultseppunct}\relax
\EndOfBibitem
\bibitem[Baiardi \latin{et~al.}(2015)Baiardi, Bloino, and
  Barone]{Baiardi15jctc}
Baiardi,~A.; Bloino,~J.; Barone,~V. Accurate {Simulation} of
  {Resonance}-{Raman} {Spectra} of {Flexible} {Molecules}: {An} {Internal}
  {Coordinates} {Approach}. \emph{Journal of Chemical Theory and Computation}
  \textbf{2015}, \emph{11}, 3267--3280\relax
\mciteBstWouldAddEndPuncttrue
\mciteSetBstMidEndSepPunct{\mcitedefaultmidpunct}
{\mcitedefaultendpunct}{\mcitedefaultseppunct}\relax
\EndOfBibitem
\bibitem[Heller \latin{et~al.}(2015)Heller, Yang, and Kocia]{Heller15}
Heller,~E.~J.; Yang,~Y.; Kocia,~L. Raman {Scattering} in {Carbon}
  {Nanosystems}: {Solving} {Polyacetylene}. \emph{ACS Central Science}
  \textbf{2015}, \emph{1}, 40--49\relax
\mciteBstWouldAddEndPuncttrue
\mciteSetBstMidEndSepPunct{\mcitedefaultmidpunct}
{\mcitedefaultendpunct}{\mcitedefaultseppunct}\relax
\EndOfBibitem
\bibitem[Kramers and Heisenberg(1925)Kramers, and Heisenberg]{Kramers25zp}
Kramers,~H.~A.; Heisenberg,~W. Über die {Streuung} von {Strahlung} durch
  {Atome}. \emph{Zeitschrift für Physik} \textbf{1925}, \emph{31},
  681--708\relax
\mciteBstWouldAddEndPuncttrue
\mciteSetBstMidEndSepPunct{\mcitedefaultmidpunct}
{\mcitedefaultendpunct}{\mcitedefaultseppunct}\relax
\EndOfBibitem
\bibitem[Dirac(1927)]{Dirac27prsla}
Dirac,~P. a.~M. The {Quantum} {Theory} of {Dispersion}. \emph{Proceedings of
  the Royal Society of London A: Mathematical, Physical and Engineering
  Sciences} \textbf{1927}, \emph{114}, 710--728\relax
\mciteBstWouldAddEndPuncttrue
\mciteSetBstMidEndSepPunct{\mcitedefaultmidpunct}
{\mcitedefaultendpunct}{\mcitedefaultseppunct}\relax
\EndOfBibitem
\bibitem[Breit(1932)]{Breit32rmp}
Breit,~G. Quantum {Theory} of {Dispersion}. \emph{Reviews of Modern Physics}
  \textbf{1932}, \emph{4}, 504--576\relax
\mciteBstWouldAddEndPuncttrue
\mciteSetBstMidEndSepPunct{\mcitedefaultmidpunct}
{\mcitedefaultendpunct}{\mcitedefaultseppunct}\relax
\EndOfBibitem
\bibitem[Jensen \latin{et~al.}(2005)Jensen, Autschbach, and Schatz]{Jensen05}
Jensen,~L.; Autschbach,~J.; Schatz,~G.~C. Finite lifetime effects on the
  polarizability within time-dependent density-functional theory. \emph{The
  Journal of Chemical Physics} \textbf{2005}, \emph{122}, 224115\relax
\mciteBstWouldAddEndPuncttrue
\mciteSetBstMidEndSepPunct{\mcitedefaultmidpunct}
{\mcitedefaultendpunct}{\mcitedefaultseppunct}\relax
\EndOfBibitem
\bibitem[Neese \latin{et~al.}(2007)Neese, Petrenko, Ganyushin, and
  Olbrich]{Neese07ccr}
Neese,~F.; Petrenko,~T.; Ganyushin,~D.; Olbrich,~G. Advanced aspects of ab
  initio theoretical optical spectroscopy of transition metal complexes:
  {Multiplets}, spin-orbit coupling and resonance {Raman} intensities.
  \emph{Coordination Chemistry Reviews} \textbf{2007}, \emph{251},
  288--327\relax
\mciteBstWouldAddEndPuncttrue
\mciteSetBstMidEndSepPunct{\mcitedefaultmidpunct}
{\mcitedefaultendpunct}{\mcitedefaultseppunct}\relax
\EndOfBibitem
\bibitem[Lee(1983)]{Lee83jcp}
Lee,~S. Placzek‐type polarizability tensors for {Raman} and resonance {Raman}
  scattering. \emph{The Journal of Chemical Physics} \textbf{1983}, \emph{78},
  723--734\relax
\mciteBstWouldAddEndPuncttrue
\mciteSetBstMidEndSepPunct{\mcitedefaultmidpunct}
{\mcitedefaultendpunct}{\mcitedefaultseppunct}\relax
\EndOfBibitem
\bibitem[Jensen \latin{et~al.}(2005)Jensen, Zhao, Autschbach, and
  Schatz]{Jensen05jcp}
Jensen,~L.; Zhao,~L.~L.; Autschbach,~J.; Schatz,~G.~C. Theory and method for
  calculating resonance {Raman} scattering from resonance polarizability
  derivatives. \emph{The Journal of Chemical Physics} \textbf{2005},
  \emph{123}, 174110\relax
\mciteBstWouldAddEndPuncttrue
\mciteSetBstMidEndSepPunct{\mcitedefaultmidpunct}
{\mcitedefaultendpunct}{\mcitedefaultseppunct}\relax
\EndOfBibitem
\bibitem[Montero(1982)]{Montero82jcp}
Montero,~S. Anharmonic {Raman} intensities of overtones, combination and
  difference bands. \emph{The Journal of Chemical Physics} \textbf{1982},
  \emph{77}, 23--29\relax
\mciteBstWouldAddEndPuncttrue
\mciteSetBstMidEndSepPunct{\mcitedefaultmidpunct}
{\mcitedefaultendpunct}{\mcitedefaultseppunct}\relax
\EndOfBibitem
\bibitem[Hemert and Blom(1981)Hemert, and Blom]{Hemert81mp}
Hemert,~M. C.~V.; Blom,~C.~E. Ab initio calculations of {Raman} intensities;
  analysis of the bond polarizability approach and the atom dipole interaction
  model. \emph{Molecular Physics} \textbf{1981}, \emph{43}, 229--250\relax
\mciteBstWouldAddEndPuncttrue
\mciteSetBstMidEndSepPunct{\mcitedefaultmidpunct}
{\mcitedefaultendpunct}{\mcitedefaultseppunct}\relax
\EndOfBibitem
\bibitem[{András Stirling}(1996)]{Stirling96jcp}
{András Stirling}, Raman intensities from {Kohn}–{Sham} calculations.
  \emph{The Journal of Chemical Physics} \textbf{1996}, \emph{104},
  1254--1262\relax
\mciteBstWouldAddEndPuncttrue
\mciteSetBstMidEndSepPunct{\mcitedefaultmidpunct}
{\mcitedefaultendpunct}{\mcitedefaultseppunct}\relax
\EndOfBibitem
\bibitem[Shinohara \latin{et~al.}(1998)Shinohara, Yamakita, and
  Ohno]{Shinohara98jms}
Shinohara,~H.; Yamakita,~Y.; Ohno,~K. Raman spectra of polycyclic aromatic
  hydrocarbons. {Comparison} of calculated {Raman} intensity distributions with
  observed spectra for naphthalene, anthracene, pyrene, and perylene.
  \emph{Journal of Molecular Structure} \textbf{1998}, \emph{442},
  221--234\relax
\mciteBstWouldAddEndPuncttrue
\mciteSetBstMidEndSepPunct{\mcitedefaultmidpunct}
{\mcitedefaultendpunct}{\mcitedefaultseppunct}\relax
\EndOfBibitem
\bibitem[Jackson \latin{et~al.}(1997)Jackson, Pederson, Porezag, Hajnal, and
  Frauenheim]{Jackson97prb}
Jackson,~K.; Pederson,~M.~R.; Porezag,~D.; Hajnal,~Z.; Frauenheim,~T.
  Density-functional-based predictions of {Raman} and {IR} spectra for small
  {Si} clusters. \emph{Physical Review B} \textbf{1997}, \emph{55},
  2549--2555\relax
\mciteBstWouldAddEndPuncttrue
\mciteSetBstMidEndSepPunct{\mcitedefaultmidpunct}
{\mcitedefaultendpunct}{\mcitedefaultseppunct}\relax
\EndOfBibitem
\bibitem[Umari and Pasquarello(2005)Umari, and Pasquarello]{Umari05drm}
Umari,~P.; Pasquarello,~A. Infrared and {Raman} spectra of disordered materials
  from first principles. \emph{Diamond and Related Materials} \textbf{2005},
  \emph{14}, 1255--1261\relax
\mciteBstWouldAddEndPuncttrue
\mciteSetBstMidEndSepPunct{\mcitedefaultmidpunct}
{\mcitedefaultendpunct}{\mcitedefaultseppunct}\relax
\EndOfBibitem
\bibitem[Ferrari and Robertson(2000)Ferrari, and Robertson]{Ferrari00prb}
Ferrari,~A.~C.; Robertson,~J. Interpretation of {Raman} spectra of disordered
  and amorphous carbon. \emph{Physical Review B} \textbf{2000}, \emph{61},
  14095--14107\relax
\mciteBstWouldAddEndPuncttrue
\mciteSetBstMidEndSepPunct{\mcitedefaultmidpunct}
{\mcitedefaultendpunct}{\mcitedefaultseppunct}\relax
\EndOfBibitem
\bibitem[Piscanec \latin{et~al.}(2005)Piscanec, Mauri, Ferrari, Lazzeri, and
  Robertson]{Piscanec05drm}
Piscanec,~S.; Mauri,~F.; Ferrari,~A.~C.; Lazzeri,~M.; Robertson,~J. Ab initio
  resonant {Raman} spectra of diamond-like carbons. \emph{Diamond and Related
  Materials} \textbf{2005}, \emph{14}, 1078--1083\relax
\mciteBstWouldAddEndPuncttrue
\mciteSetBstMidEndSepPunct{\mcitedefaultmidpunct}
{\mcitedefaultendpunct}{\mcitedefaultseppunct}\relax
\EndOfBibitem
\bibitem[Rousseau and Williams(1976)Rousseau, and Williams]{Rousseau76}
Rousseau,~D.~L.; Williams,~P.~F. Resonance {Raman} scattering of light from a
  diatomic molecule. \emph{The Journal of Chemical Physics} \textbf{1976},
  \emph{64}, 3519--3537\relax
\mciteBstWouldAddEndPuncttrue
\mciteSetBstMidEndSepPunct{\mcitedefaultmidpunct}
{\mcitedefaultendpunct}{\mcitedefaultseppunct}\relax
\EndOfBibitem
\bibitem[Dierksen and Grimme(2004)Dierksen, and Grimme]{Dierksen04jcp}
Dierksen,~M.; Grimme,~S. Density functional calculations of the vibronic
  structure of electronic absorption spectra. \emph{The Journal of Chemical
  Physics} \textbf{2004}, \emph{120}, 3544--3554\relax
\mciteBstWouldAddEndPuncttrue
\mciteSetBstMidEndSepPunct{\mcitedefaultmidpunct}
{\mcitedefaultendpunct}{\mcitedefaultseppunct}\relax
\EndOfBibitem
\bibitem[Moran \latin{et~al.}(2002)Moran, Egolf, Blanchard-Desce, and
  Kelley]{Moran02jcp}
Moran,~A.~M.; Egolf,~D.~S.; Blanchard-Desce,~M.; Kelley,~A.~M. Vibronic effects
  on solvent dependent linear and nonlinear optical properties of push-pull
  chromophores: {Julolidinemalononitrile}. \emph{The Journal of Chemical
  Physics} \textbf{2002}, \emph{116}, 2542--2555\relax
\mciteBstWouldAddEndPuncttrue
\mciteSetBstMidEndSepPunct{\mcitedefaultmidpunct}
{\mcitedefaultendpunct}{\mcitedefaultseppunct}\relax
\EndOfBibitem
\bibitem[Jarzecki(2009)]{Jarzecki09jpca}
Jarzecki,~A.~A. Quantum-{Mechanical} {Calculations} of {Resonance} {Raman}
  {Intensities}: {The} {Weighted}-{Gradient} {Approximation}. \emph{The Journal
  of Physical Chemistry A} \textbf{2009}, \emph{113}, 2926--2934\relax
\mciteBstWouldAddEndPuncttrue
\mciteSetBstMidEndSepPunct{\mcitedefaultmidpunct}
{\mcitedefaultendpunct}{\mcitedefaultseppunct}\relax
\EndOfBibitem
\bibitem[Wächtler \latin{et~al.}(2012)Wächtler, Guthmuller, González, and
  Dietzek]{Wachtler12}
Wächtler,~M.; Guthmuller,~J.; González,~L.; Dietzek,~B. Analysis and
  characterization of coordination compounds by resonance {Raman} spectroscopy.
  \emph{Coordination Chemistry Reviews} \textbf{2012}, \emph{256},
  1479--1508\relax
\mciteBstWouldAddEndPuncttrue
\mciteSetBstMidEndSepPunct{\mcitedefaultmidpunct}
{\mcitedefaultendpunct}{\mcitedefaultseppunct}\relax
\EndOfBibitem
\bibitem[Korenowski \latin{et~al.}(1978)Korenowski, Ziegler, and
  Albrecht]{Korenowski78jcp}
Korenowski,~G.~M.; Ziegler,~L.~D.; Albrecht,~A.~C. Calculations of resonance
  {Raman} cross sections in forbidden electronic transitions: {Scattering} of
  the 992 cm$^{-1}$ mode in the 1B2u band of benzene. \emph{The Journal of
  Chemical Physics} \textbf{1978}, \emph{68}, 1248--1252\relax
\mciteBstWouldAddEndPuncttrue
\mciteSetBstMidEndSepPunct{\mcitedefaultmidpunct}
{\mcitedefaultendpunct}{\mcitedefaultseppunct}\relax
\EndOfBibitem
\bibitem[Woodward and Long(1949)Woodward, and Long]{Woodward49tfs}
Woodward,~L.~A.; Long,~D.~A. Relative intensities in the {Raman} spectra of
  some {Group} {IV} tetrahalides. \emph{Transactions of the Faraday Society}
  \textbf{1949}, \emph{45}, 1131--1141\relax
\mciteBstWouldAddEndPuncttrue
\mciteSetBstMidEndSepPunct{\mcitedefaultmidpunct}
{\mcitedefaultendpunct}{\mcitedefaultseppunct}\relax
\EndOfBibitem
\bibitem[Mortensen \latin{et~al.}(2005)Mortensen, Hansen, and
  Jacobsen]{Mortensen05prb}
Mortensen,~J.~J.; Hansen,~L.~B.; Jacobsen,~K.~W. Real-space grid implementation
  of the projector augmented wave method. \emph{Physical Review B}
  \textbf{2005}, \emph{71}, 035109\relax
\mciteBstWouldAddEndPuncttrue
\mciteSetBstMidEndSepPunct{\mcitedefaultmidpunct}
{\mcitedefaultendpunct}{\mcitedefaultseppunct}\relax
\EndOfBibitem
\bibitem[Enkovaara \latin{et~al.}(2010)Enkovaara, Rostgaard, Mortensen, Chen,
  Dułak, Ferrighi, Gavnholt, Glinsvad, Haikola, Hansen, Kristoffersen, Kuisma,
  Larsen, Lehtovaara, Ljungberg, Lopez-Acevedo, Moses, Ojanen, Olsen, Petzold,
  Romero, Stausholm-Møller, Strange, Tritsaris, Vanin, Walter, Hammer,
  Häkkinen, Madsen, Nieminen, Nørskov, Puska, Rantala, Schiøtz, Thygesen,
  and Jacobsen]{Enkovaara10jpc}
Enkovaara,~J.; Rostgaard,~C.; Mortensen,~J.~J.; Chen,~J.; Dułak,~M.;
  Ferrighi,~L.; Gavnholt,~J.; Glinsvad,~C.; Haikola,~V.; Hansen,~H.~A.;
  Kristoffersen,~H.~H.; Kuisma,~M.; Larsen,~A.~H.; Lehtovaara,~L.;
  Ljungberg,~M.; Lopez-Acevedo,~O.; Moses,~P.~G.; Ojanen,~J.; Olsen,~T.;
  Petzold,~V.; Romero,~N.~A.; Stausholm-Møller,~J.; Strange,~M.;
  Tritsaris,~G.~A.; Vanin,~M.; Walter,~M.; Hammer,~B.; Häkkinen,~H.;
  Madsen,~G. K.~H.; Nieminen,~R.~M.; Nørskov,~J.~K.; Puska,~M.;
  Rantala,~T.~T.; Schiøtz,~J.; Thygesen,~K.~S.; Jacobsen,~K.~W. Electronic
  structure calculations with {GPAW}: a real-space implementation of the
  projector augmented-wave method. \emph{Journal of Physics: Condensed Matter}
  \textbf{2010}, \emph{22}, 253202\relax
\mciteBstWouldAddEndPuncttrue
\mciteSetBstMidEndSepPunct{\mcitedefaultmidpunct}
{\mcitedefaultendpunct}{\mcitedefaultseppunct}\relax
\EndOfBibitem
\bibitem[Blöchl(1994)]{Blochl94prb}
Blöchl,~P.~E. Projector augmented-wave method. \emph{Physical Review B}
  \textbf{1994}, \emph{50}, 17953--17979\relax
\mciteBstWouldAddEndPuncttrue
\mciteSetBstMidEndSepPunct{\mcitedefaultmidpunct}
{\mcitedefaultendpunct}{\mcitedefaultseppunct}\relax
\EndOfBibitem
\bibitem[Perdew \latin{et~al.}(1996)Perdew, Burke, and Ernzerhof]{Perdew96prl}
Perdew,~J.~P.; Burke,~K.; Ernzerhof,~M. Generalized {Gradient} {Approximation}
  {Made} {Simple}. \emph{Physical Review Letters} \textbf{1996}, \emph{77},
  3865--3868\relax
\mciteBstWouldAddEndPuncttrue
\mciteSetBstMidEndSepPunct{\mcitedefaultmidpunct}
{\mcitedefaultendpunct}{\mcitedefaultseppunct}\relax
\EndOfBibitem
\bibitem[Larsen \latin{et~al.}(2017)Larsen, Mortensen, Blomqvist, Castelli,
  Christensen, {Marcin Dułak}, Friis, Groves, Hammer, Hargus, Hermes,
  Jennings, Jensen, Kermode, Kitchin, Kolsbjerg, Kubal, {Kristen Kaasbjerg},
  Lysgaard, Maronsson, Maxson, Olsen, Pastewka, {Andrew Peterson}, Rostgaard,
  Schiøtz, Schütt, Strange, Thygesen, {Tejs Vegge}, Vilhelmsen, Walter, Zeng,
  and Jacobsen]{Larsen17jpc}
Larsen,~A.~H.; Mortensen,~J.~J.; Blomqvist,~J.; Castelli,~I.~E.;
  Christensen,~R.; {Marcin Dułak},; Friis,~J.; Groves,~M.~N.; Hammer,~B.;
  Hargus,~C.; Hermes,~E.~D.; Jennings,~P.~C.; Jensen,~P.~B.; Kermode,~J.;
  Kitchin,~J.~R.; Kolsbjerg,~E.~L.; Kubal,~J.; {Kristen Kaasbjerg},;
  Lysgaard,~S.; Maronsson,~J.~B.; Maxson,~T.; Olsen,~T.; Pastewka,~L.; {Andrew
  Peterson},; Rostgaard,~C.; Schiøtz,~J.; Schütt,~O.; Strange,~M.;
  Thygesen,~K.~S.; {Tejs Vegge},; Vilhelmsen,~L.; Walter,~M.; Zeng,~Z.;
  Jacobsen,~K.~W. The atomic simulation environment—a {Python} library for
  working with atoms. \emph{Journal of Physics: Condensed Matter}
  \textbf{2017}, \emph{29}, 273002\relax
\mciteBstWouldAddEndPuncttrue
\mciteSetBstMidEndSepPunct{\mcitedefaultmidpunct}
{\mcitedefaultendpunct}{\mcitedefaultseppunct}\relax
\EndOfBibitem
\bibitem[{M.E. Casida} and {M. Huix-Rotllant}(2012){M.E. Casida}, and {M.
  Huix-Rotllant}]{Casida12arpc}
{M.E. Casida},; {M. Huix-Rotllant}, Progress in {Time}-{Dependent}
  {Density}-{Functional} {Theory} {\textbar} {Annual} {Review} of {Physical}
  {Chemistry}. \emph{Annual Review of Physical Chemistry} \textbf{2012},
  \emph{63}, 287--323\relax
\mciteBstWouldAddEndPuncttrue
\mciteSetBstMidEndSepPunct{\mcitedefaultmidpunct}
{\mcitedefaultendpunct}{\mcitedefaultseppunct}\relax
\EndOfBibitem
\bibitem[Walter \latin{et~al.}(2008)Walter, Häkkinen, Lehtovaara, Puska,
  Enkovaara, Rostgaard, and Mortensen]{Walter08jcp}
Walter,~M.; Häkkinen,~H.; Lehtovaara,~L.; Puska,~M.; Enkovaara,~J.;
  Rostgaard,~C.; Mortensen,~J.~J. Time-dependent density-functional theory in
  the projector augmented-wave method. \emph{The Journal of Chemical Physics}
  \textbf{2008}, \emph{128}, 244101\relax
\mciteBstWouldAddEndPuncttrue
\mciteSetBstMidEndSepPunct{\mcitedefaultmidpunct}
{\mcitedefaultendpunct}{\mcitedefaultseppunct}\relax
\EndOfBibitem
\bibitem[Oklopčić \latin{et~al.}(2016)Oklopčić, Hirata, and
  Heng]{Oklopcic16}
Oklopčić,~A.; Hirata,~C.~M.; Heng,~K. Raman {Scattering} by {Molecular}
  {Hydrogen} and {Nitrogen} in {Exoplanetary} {Atmospheres}. \emph{The
  Astrophysical Journal} \textbf{2016}, \emph{832}, 30\relax
\mciteBstWouldAddEndPuncttrue
\mciteSetBstMidEndSepPunct{\mcitedefaultmidpunct}
{\mcitedefaultendpunct}{\mcitedefaultseppunct}\relax
\EndOfBibitem
\bibitem[Wolniewicz and Staszewska(2003)Wolniewicz, and
  Staszewska]{Wolniewicz03}
Wolniewicz,~L.; Staszewska,~G. 1{$\Sigma$u}+$\to${X}1{$\Sigma$g}+ transition
  moments for the hydrogen molecule. \emph{Journal of Molecular Spectroscopy}
  \textbf{2003}, \emph{217}, 181--185\relax
\mciteBstWouldAddEndPuncttrue
\mciteSetBstMidEndSepPunct{\mcitedefaultmidpunct}
{\mcitedefaultendpunct}{\mcitedefaultseppunct}\relax
\EndOfBibitem
\bibitem[Fantz and Wünderlich(2006)Fantz, and Wünderlich]{Fantz06}
Fantz,~U.; Wünderlich,~D. Franck–{Condon} factors, transition probabilities,
  and radiative lifetimes for hydrogen molecules and their isotopomeres.
  \emph{Atomic Data and Nuclear Data Tables} \textbf{2006}, \emph{92},
  853--973\relax
\mciteBstWouldAddEndPuncttrue
\mciteSetBstMidEndSepPunct{\mcitedefaultmidpunct}
{\mcitedefaultendpunct}{\mcitedefaultseppunct}\relax
\EndOfBibitem
\bibitem[Jong \latin{et~al.}(2015)Jong, Seijo, Meijerink, and
  Rabouw]{Jong15pccp}
Jong,~M.~d.; Seijo,~L.; Meijerink,~A.; Rabouw,~F.~T. Resolving the ambiguity in
  the relation between {Stokes} shift and {Huang}–{Rhys} parameter.
  \emph{Physical Chemistry Chemical Physics} \textbf{2015}, \emph{17},
  16959--16969\relax
\mciteBstWouldAddEndPuncttrue
\mciteSetBstMidEndSepPunct{\mcitedefaultmidpunct}
{\mcitedefaultendpunct}{\mcitedefaultseppunct}\relax
\EndOfBibitem
\bibitem[{W. S. Benedict} \latin{et~al.}(1956){W. S. Benedict}, {N. Gailar},
  and {Earle K. Plyler}]{Benedict56jcp}
{W. S. Benedict},; {N. Gailar},; {Earle K. Plyler}, Rotation‐{Vibration}
  {Spectra} of {Deuterated} {Water} {Vapor}. \emph{The Journal of Chemical
  Physics} \textbf{1956}, \emph{24}, 1139--1165\relax
\mciteBstWouldAddEndPuncttrue
\mciteSetBstMidEndSepPunct{\mcitedefaultmidpunct}
{\mcitedefaultendpunct}{\mcitedefaultseppunct}\relax
\EndOfBibitem
\bibitem[{Benny G. Johnson} \latin{et~al.}(1993){Benny G. Johnson}, {Peter M.
  W. Gill}, and {John A. Pople}]{Johnson93jcp}
{Benny G. Johnson},; {Peter M. W. Gill},; {John A. Pople}, The performance of a
  family of density functional methods. \emph{The Journal of Chemical Physics}
  \textbf{1993}, \emph{98}, 5612--5626\relax
\mciteBstWouldAddEndPuncttrue
\mciteSetBstMidEndSepPunct{\mcitedefaultmidpunct}
{\mcitedefaultendpunct}{\mcitedefaultseppunct}\relax
\EndOfBibitem
\bibitem[{Dimitrij Rappoport}(2007)]{Rappoport07}
{Dimitrij Rappoport}, \emph{Berechnung von {Raman}-{Intensitäten} mit
  zeitabhängiger {Dichtefunktionaltheorie}}; Univ.-Verl. Karlsruhe: Karlsruhe,
  2007\relax
\mciteBstWouldAddEndPuncttrue
\mciteSetBstMidEndSepPunct{\mcitedefaultmidpunct}
{\mcitedefaultendpunct}{\mcitedefaultseppunct}\relax
\EndOfBibitem
\bibitem[Caillie and Amos(2000)Caillie, and Amos]{Caillie00pccp}
Caillie,~C.~V.; Amos,~R.~D. Raman intensities using time dependent density
  functional theory. \emph{Physical Chemistry Chemical Physics} \textbf{2000},
  \emph{2}, 2123--2129\relax
\mciteBstWouldAddEndPuncttrue
\mciteSetBstMidEndSepPunct{\mcitedefaultmidpunct}
{\mcitedefaultendpunct}{\mcitedefaultseppunct}\relax
\EndOfBibitem
\bibitem[Richards and Nielsen(1950)Richards, and Nielsen]{Richards50}
Richards,~C.~M.; Nielsen,~J.~R. Raman {Spectrum} of 1,3-{Butadiene} in the
  {Gaseous} and {Liquid} {States}*. \emph{JOSA} \textbf{1950}, \emph{40},
  438--441\relax
\mciteBstWouldAddEndPuncttrue
\mciteSetBstMidEndSepPunct{\mcitedefaultmidpunct}
{\mcitedefaultendpunct}{\mcitedefaultseppunct}\relax
\EndOfBibitem
\bibitem[Chadwick \latin{et~al.}(1991)Chadwick, Zgierski, and
  Hudson]{Chadwick91jcp}
Chadwick,~R.~R.; Zgierski,~M.~Z.; Hudson,~B.~S. Resonance {Raman} scattering of
  butadiene: {Vibronic} activity of a bu mode demonstrates the presence of a
  1Ag symmetry excited electronic state at low energy. \emph{The Journal of
  Chemical Physics} \textbf{1991}, \emph{95}, 7204--7211\relax
\mciteBstWouldAddEndPuncttrue
\mciteSetBstMidEndSepPunct{\mcitedefaultmidpunct}
{\mcitedefaultendpunct}{\mcitedefaultseppunct}\relax
\EndOfBibitem
\bibitem[Phillips \latin{et~al.}(1993)Phillips, Zgierski, and
  Myers]{Phillips93jpc}
Phillips,~D.~L.; Zgierski,~M.~Z.; Myers,~A.~B. Resonance {Raman} excitation
  profiles of 1,3-butadiene in vapor and solution phases. \emph{The Journal of
  Physical Chemistry} \textbf{1993}, \emph{97}, 1800--1809\relax
\mciteBstWouldAddEndPuncttrue
\mciteSetBstMidEndSepPunct{\mcitedefaultmidpunct}
{\mcitedefaultendpunct}{\mcitedefaultseppunct}\relax
\EndOfBibitem
\bibitem[Krawczyk \latin{et~al.}(2000)Krawczyk, Malsch, Hohlneicher, Gillen,
  and Domcke]{Krawczyk00cpl}
Krawczyk,~R.~P.; Malsch,~K.; Hohlneicher,~G.; Gillen,~R.~C.; Domcke,~W. 1
  1Bu–2 1Ag conical intersection in trans-butadiene: ultrafast dynamics and
  optical spectra. \emph{Chemical Physics Letters} \textbf{2000}, \emph{320},
  535--541\relax
\mciteBstWouldAddEndPuncttrue
\mciteSetBstMidEndSepPunct{\mcitedefaultmidpunct}
{\mcitedefaultendpunct}{\mcitedefaultseppunct}\relax
\EndOfBibitem
\bibitem[Peng \latin{et~al.}(2007)Peng, Yi, Shuai, and Shao]{Peng07jacs}
Peng,~Q.; Yi,~Y.; Shuai,~Z.; Shao,~J. Toward {Quantitative} {Prediction} of
  {Molecular} {Fluorescence} {Quantum} {Efficiency}: {Role} of {Duschinsky}
  {Rotation}. \emph{Journal of the American Chemical Society} \textbf{2007},
  \emph{129}, 9333--9339\relax
\mciteBstWouldAddEndPuncttrue
\mciteSetBstMidEndSepPunct{\mcitedefaultmidpunct}
{\mcitedefaultendpunct}{\mcitedefaultseppunct}\relax
\EndOfBibitem
\bibitem[Hemley \latin{et~al.}(1983)Hemley, Dawson, and Vaida]{Hemley83jcp}
Hemley,~R.~J.; Dawson,~J.~I.; Vaida,~V. Franck–{Condon} analysis of the $1
  ^1A_g^-\to ^1B_u^+$ transition of 1,3‐butadiene from absorption and {Raman}
  intensities. \emph{The Journal of Chemical Physics} \textbf{1983}, \emph{78},
  2915--2927\relax
\mciteBstWouldAddEndPuncttrue
\mciteSetBstMidEndSepPunct{\mcitedefaultmidpunct}
{\mcitedefaultendpunct}{\mcitedefaultseppunct}\relax
\EndOfBibitem
\bibitem[Hsu \latin{et~al.}(2001)Hsu, Hirata, and Head-Gordon]{Hsu01jpca}
Hsu,~C.-P.; Hirata,~S.; Head-Gordon,~M. Excitation {Energies} from
  {Time}-{Dependent} {Density} {Functional} {Theory} for {Linear} {Polyene}
  {Oligomers: Butadiene} to {Decapentaene}. \emph{The Journal of Physical
  Chemistry A} \textbf{2001}, \emph{105}, 451--458\relax
\mciteBstWouldAddEndPuncttrue
\mciteSetBstMidEndSepPunct{\mcitedefaultmidpunct}
{\mcitedefaultendpunct}{\mcitedefaultseppunct}\relax
\EndOfBibitem
\bibitem[Hsu \latin{et~al.}(2001)Hsu, Hirata, and Head-Gordon]{Hsu01jcpa}
Hsu,~C.-P.; Hirata,~S.; Head-Gordon,~M. Excitation {Energies} from
  {Time}-{Dependent} {Density} {Functional} {Theory} for {Linear} {Polyene}
  {Oligomers: Butadiene} to {Decapentaene}. \emph{The Journal of Physical
  Chemistry A} \textbf{2001}, \emph{105}, 451--458\relax
\mciteBstWouldAddEndPuncttrue
\mciteSetBstMidEndSepPunct{\mcitedefaultmidpunct}
{\mcitedefaultendpunct}{\mcitedefaultseppunct}\relax
\EndOfBibitem
\bibitem[{Stefan Grimme}(2004)]{Grimme04}
{Stefan Grimme}, \emph{Reviews in {Computational} {Chemistry}, {Volume} 20};
  John Wiley \& Sons, Inc, 2004; Vol.~20; p 153\relax
\mciteBstWouldAddEndPuncttrue
\mciteSetBstMidEndSepPunct{\mcitedefaultmidpunct}
{\mcitedefaultendpunct}{\mcitedefaultseppunct}\relax
\EndOfBibitem
\bibitem[Guo \latin{et~al.}(2012)Guo, He, Dai, Shen, Li, Zhu, and Lin]{Guo12}
Guo,~M.; He,~R.; Dai,~Y.; Shen,~W.; Li,~M.; Zhu,~C.; Lin,~S.~H. Franck-{Condon}
  simulation of vibrationally resolved optical spectra for zinc complexes of
  phthalocyanine and tetrabenzoporphyrin including the {Duschinsky} and
  {Herzberg}-{Teller} effects. \emph{The Journal of Chemical Physics}
  \textbf{2012}, \emph{136}, 144313\relax
\mciteBstWouldAddEndPuncttrue
\mciteSetBstMidEndSepPunct{\mcitedefaultmidpunct}
{\mcitedefaultendpunct}{\mcitedefaultseppunct}\relax
\EndOfBibitem
\bibitem[Keil(1965)]{Keil65}
Keil,~T.~H. Shapes of {Impurity} {Absorption} {Bands} in {Solids}.
  \emph{Physical Review} \textbf{1965}, \emph{140}, A601--A617\relax
\mciteBstWouldAddEndPuncttrue
\mciteSetBstMidEndSepPunct{\mcitedefaultmidpunct}
{\mcitedefaultendpunct}{\mcitedefaultseppunct}\relax
\EndOfBibitem
\bibitem[Resta(2000)]{Resta00jpc}
Resta,~R. Manifestations of {Berry}'s phase in molecules and condensed matter.
  \emph{Journal of Physics: Condensed Matter} \textbf{2000}, \emph{12},
  R107\relax
\mciteBstWouldAddEndPuncttrue
\mciteSetBstMidEndSepPunct{\mcitedefaultmidpunct}
{\mcitedefaultendpunct}{\mcitedefaultseppunct}\relax
\EndOfBibitem
\bibitem[Min \latin{et~al.}(2014)Min, Abedi, Kim, and Gross]{Min14prl}
Min,~S.~K.; Abedi,~A.; Kim,~K.~S.; Gross,~E. Is the {Molecular} {Berry} {Phase}
  an {Artifact} of the {Born}-{Oppenheimer} {Approximation}? \emph{Physical
  Review Letters} \textbf{2014}, \emph{113}, 263004\relax
\mciteBstWouldAddEndPuncttrue
\mciteSetBstMidEndSepPunct{\mcitedefaultmidpunct}
{\mcitedefaultendpunct}{\mcitedefaultseppunct}\relax
\EndOfBibitem
\bibitem[Baer(2002)]{Baer02jcp}
Baer,~R. Born–{Oppenheimer} invariants along nuclear configuration paths.
  \emph{The Journal of Chemical Physics} \textbf{2002}, \emph{117},
  7405--7408\relax
\mciteBstWouldAddEndPuncttrue
\mciteSetBstMidEndSepPunct{\mcitedefaultmidpunct}
{\mcitedefaultendpunct}{\mcitedefaultseppunct}\relax
\EndOfBibitem
\bibitem[Casida(2009)]{Casida09}
Casida,~M.~E. Time-Dependent Density-Functional Theory for Molecules and
  Molecular Solids. \emph{Journal of Molecular Structure: THEOCHEM}
  \textbf{2009}, \emph{914}, 3--18\relax
\mciteBstWouldAddEndPuncttrue
\mciteSetBstMidEndSepPunct{\mcitedefaultmidpunct}
{\mcitedefaultendpunct}{\mcitedefaultseppunct}\relax
\EndOfBibitem
\end{mcitethebibliography}
\bibliographystyle{apsrev4-1}

\end{document}